\documentclass[prb,twocolumn,aps,superscriptaddress,nofootinbib,longbibliography,10pt]{revtex4-2}
\usepackage[ansinew]{inputenc}
\usepackage[caption=false]{subfig}
\usepackage{bm,color,soul,amsmath,amssymb,mathrsfs,latexsym,graphicx,booktabs,array,psfrag,float}

\DeclareMathOperator{\cosec}{cosec}
\usepackage[english]{babel}
\DeclareMathOperator{\sgn}{sgn}

\begin{document}
\title{\Large\textbf{Anisotropic non-linear optical response of nodal loop materials}}
\author{Tommy Tai}
\affiliation{Cavendish Laboratory, University of Cambridge, JJ Thomson Ave, Cambridge CB3 0HE, United Kingdom}
\email{ytt26@cam.ac.uk}
\author{Ching Hua Lee}
\affiliation{Department of Physics, National University of Singapore, Singapore 117542}
\email{phylch@nus.edu.sg}
\date{\today}

\setlength{\parindent}{0cm}
\begin{abstract}
Nodal line semimetals have lately aroused much experimental and theoretical interest, with their gap closing along unconventional trajectories in the 3D Brillouin zone. These trajectories or nodal lines can close into loops and trace out intricate knotted or linked configurations with complicated topologies. In this work, we investigate the semi-classical optical response of two nodal loops in linked, unlinked and touching configurations, focusing particularly on the interplay of response anisotropy and non-linearity. We provide a geometric picture that unifies these aspects and sheds light on the effects of nodal topology and geometry. Based on a model abstracted from generic multi nodal loop scenarios, both with or without linkages, our findings will be applicable for a large class of nodal semimetal materials with multiple nodal lines or loops.
\end{abstract}
\maketitle

\section{Introduction }
Topological materials are among the most intensely researched current topics, and encompass not just topological insulators~\cite{kane2005quantum,kane2005z,PhysRevLett.96.106802,bernevig2006quantum, fu2007topological,PhysRevLett.106.106802, hsieh2012topological,tanaka2012experimental,dziawa2012topological,PhysRevB.78.195125}
but also semi-metals with topological nontrivial nodal structures~\cite{burkov2016topological,doi:10.1146/annurev-matsci-070218-010049, doi:10.7566/JPSJ.87.041001, fang2016topological,doi:10.1146/annurev-matsci-070218-010023}. Such nodal materials have provoked widespread studies due to their unconventional density of states and bandstructure, which have lead to new or enhanced avenues for non-linear optical responses, electron tunneling behavior, higher harmonic generation, superconductivity and quantum Hall effects~\cite{RevModPhys.82.3045,doi:10.1146/annurev-conmatphys-031214-014501,RevModPhys.91.015006, ozawa2019topological, doi:10.7566/JPSJ.87.041001,dong2010quantum,pang2015evidence,wang20173d,molina2018surface,zhang2019cyclotron,zhang2019quantum,ang2020universal,bhattacharyya2020evidence}. The potentially intricate topology of nodal loops in 3D space have also inspired their design and realization in metamaterials and lossy or non-reciprocal media~\cite{lu2013weyl,xiao2015synthetic,lin2016photonic,yang2016acoustic,chen2016photonic,chang2017multiple,noh2017experimental,lin2017line,li2018weyl,yang2018ideal,gao2018experimental,lee2018tidal,xia2019observation,zhou2019exceptional,lee2020imaging,cerjan2019experimental,li2019emergence,jia2019observation}.

Despite their seemingly exotic nature, various Nodal line semimetal (NLSM) materials have recently been experimentally characterized, such as PbTaSe$_2$~\cite{bian2016topological, PhysRevB.93.245130}, BiTeI~\cite{crepaldi2012giant}, Mg$_3$Bi$_2$~\cite{doi:10.1021/acs.jpclett.7b02129,chang2019realization,zhou2019experimental}, ZrSiTe and ZrSiSe~\cite{hu2016evidence,PhysRevX.10.011026}, ZrSiS~\cite{schoop2016dirac,Fueaau6459,PhysRevB.100.085137}, BaAgAs~\cite{mardanya2019prediction}, TaN~\cite{PhysRevB.93.241202},  Ca$_3$P$_2$~\cite{chan20163,xie2015potential},  SrAs$_3$~\cite{song2020photoemission,li2018evidence}, CaAgX (X=P,As)~\cite{yamakage2016line,takane2018observation}, ZrGeX$_\text{c}$ (X$_\text{c}$= S, Se, Te) family~\cite{nakamura2019evidence, guo2019electronic}, magnetic semimetals EuB$_6$~\cite{PhysRevLett.124.076403}, spin gapless semimetals~\cite{zhang2020nodal}, the centrosymmetric superconductor SnTaS$_2$ ~\cite{chen2019superconducting} and  PbTaS$_2$~\cite{gao2020superconducting}. In the form of closed loops, they have also been observed as nodal links in CaAuAs~\cite{PhysRevB.98.085122}, nodal chains in TiB$_2$ ~\cite{yi2018observation, liu2018experimental} and nodal line networks in RuO$_2$~\cite{jovic2019dirac}. Other materials such as Ti$_3$X (X=Al, Ga, Sn, Pb) family~\cite{PhysRevB.97.235150}, YH$_3$\cite{shao2018nonsymmorphic}, YoCoC$_2$~\cite{xu2019topological}, MnN~\cite{PhysRevMaterials.3.084201}, TiTaSe$_2$~\cite{bian2016drumhead}, CaP$_3$ family~\cite{PhysRevB.95.045136}, ABC-stacked graphdyiyne~\cite{graphdiyne}, layered X$_2$Y (X=Ca, Sr, Ba; Y=As, Sb, Bi)~\cite{niu2017two} have also been theorically proposed to possess
nodal loops.

In some of these NLSMs, the conduction and the valence bands intersect to form potentially very intricate structures, even more intricate than common examples in knot theory and prototypical nodal knot setups~\cite{AlexanderPolynomial,vortexKnot,dennis2017constructing,lee2020imaging,bode2019constructing}. For instance, the material Co$_2$MnGa was theorized~\cite{CMG_theory} to possess a complicated network of 3D band crossings characterized by various types of non-trivial nodal linkages and coupled chains enabled by perpendicular mirror planes. This was subsequently confirmed experimentally~\cite{belopolski2019discovery}. Another proposal of interlocking nodal chains may be realized in carbon networks, consisting of armchair graphene nanoribbon~\cite{LI2020563}. While transport and optical response properties are already well-studied for simpler NLSMs ~\cite{doi:10.1146/annurev-matsci-070218-010023,PhysRevLett.124.076403,PhysRevB.100.085137,yang2019conductivity,guo2019electronic, PhysRevB.99.241102,PhysRevB.95.245113,PhysRevB.102.035164}, those of more complicated nodal \emph{loop} semimetals are still not well understood. Recently, it was proposed that topological nodal linkages can significantly enhance optical response non-linearity and hence HHG (higher harmonics generation)~\cite{PhysRevB.102.035138}. However, the non-linear response does not afford any topological quantization, unlike linear response via the Kubo formula, and a complete understanding of the response properties of realistic nodal loop material necessitates a systematic study of how the geometry of the nodal structure, together with its topology, interplay and lead to various anisotropic and non-linear behavior. 

As such, this work shall be concerned with a systematic investigation on how the relative shapes and linkage of nodal loops can lead to various anisotropic and non-linear responses components, and how the various responses come together into a bigger picture that reveals the overall nodal topology and geometry. Following a review of semi-classical response theory in Sect. II, we introduce a canonical model of two nodal loops whose linkage and shapes can be independently tuned. This model serves as an abstraction of the multiple nodal touchings and linkages in realistic nodal materials, for instance Co$_2$MnGa. In Sect. III, we study the various components in the response tensor of our model system, some whose non-linear properties have never been investigated. In Sect. IV, we show how these results help piece together a response surface whose evolution with field strength encapsulates the full information about the response anisotropy and non-linearity, and to some extent the nodal structure and its dispersion. For our model, the direction which the response surface points towards in the high field limit depends on whether the nodal structure is topologically linked. 

\section{Non-linear Semi-Classical Optical Response}
To systematically relate the Hopf link response to its detailed momentum-space profile, we first review the theory of non-linear semi-classical response, and next introduce our model with tunable nodal linkage. Similar semi-classical approaches have been highly successful in explaining phenomena like Hall effects and quantum oscillations in diverse settings~\cite{xiao2010berry,price2016measurement,petrides2018six,lee2018electromagnetic,lohse2018exploring}, as well as Bloch oscillations and Berry curvature effects in the context of HHG~\cite{schubert2014sub,liu2017high}. HHG refers to the high harmonics generation induced by strong non-linear interactions when a very intense laser pulse is focused into a sample.

\subsection{Semi-Classical Response Theory}
The non-linear response of nodal materials emerges even at the semi-classical level, where many-body processes are simply encapsulated by a non-equilibrium occupation function $F(\varepsilon(\mathbf{k}-e\mathbf{A(t)}))$ that weighs different contributions to the response optical current~\cite{mikhailov2007non}:
\begin{equation}
\mathbf{J}=\int F(\varepsilon(\mathbf{k}-e\mathbf{A(t)}))\langle\mathbf{k}|\mathbf{\hat J}|\mathbf{k}\rangle d^3\mathbf{k}.
\label{J}
\end{equation}
Here, $F(\varepsilon)=(1+e^{\beta(\varepsilon(\mathbf{k})-\mu)})^{-1}$ is the \emph{equilibrium} Fermi-Dirac occupation function that depends on the energy dispersion $\varepsilon(\mathbf{k})$, and $e\mathbf{A}(t)=e\int_{-\infty}^t\mathbf{E}(t')dt'$ is the impulse on an electron $e$ due to an external electric field $\mathbf{E}(t)$. In frequency space, we can always replace $\bold A$ with $\bold E/i\Omega$. The current operator for a given Hamiltonian $\hat H$ is $\mathbf{\hat J}=\frac{\partial\hat H}{\partial\mathbf{k}}$, which reduces to the dispersion velocity $\langle\mathbf{k}|\mathbf{\hat J}|\mathbf{k}\rangle=\bold v(\mathbf{k})=\frac{\partial \varepsilon(\mathbf k)}{\partial\mathbf{k}}$ in the translation-invariant intra-band case. We express this optical response in material-dependent units of $nev_F\sim10^{12}$ Am$^{-2}$ where $n$ and $v_F$ are the charge carrier density and the Fermi velocity. In real materials, we have $v_F$ to be $2.22\times10^5$ ms$^{-1}$ and $4.0\times10^5$ ms$^{-1}$ in ZrGeSe ~\cite{doi:10.1063/1.5084090} and CaAgAs~\cite{PhysRevB.95.245113} respectively, $n$ to be $3.37\times10^{26}$ m$^{-3}$, $2.98\times10^{26}$ m$^{-3}$ and $1.7\times10^{26}$ m$^{-3}$ in YbCdGe~\cite{PhysRevB.99.241102}, CaCdSn~\cite{PhysRevB.102.035164} and CaAgAs~\cite{PhysRevB.95.245113} respectively. In these NLSMs, the Fermi surface is a simple nodal loop. The optical response of real materials with more sophisticated nodal structure, beyond the simplest nodal loop, have also been computed in Ref. \cite{PhysRevB.102.035138}.

Eq.~\ref{J} holds in the ballistic limit, where it is the exact solution to the semi-classical Boltzmann equation and the semi-classical equations of motions for electronic wavepackets~\cite{mikhailov2007non}. This requires $\Omega\tau\gg 1$, where $\Omega$ is the optical frequency defined by $\mathbf{E}(t)\sim\mathbf{E}(0)e^{i\Omega t}$, and $\tau$ is the relaxation time due to impurity scattering.  For a nodal material with relaxation time $\tau\sim 10^{-12}$ to $10^{-13}\,s$, comparable to that in high quality Graphene samples~\cite{TransportTimeGraphene}, $\Omega\tau\gg 1$ can be achieved in the teraHertz regime of $\Omega \sim 50$-$100\,$THz. The relaxation times can be computed from a microscopic model for the scattering processes~\cite{ho2018theoretical}. In this ballistic regime, scattering processes cannot catch up with the much shorter oscillation timescales~\cite{liu2017high,ndabashimiye2016solid}.

According to Eq.~\ref{J}, the response current arises from contributions at $\bold k$ for which the Fermi-Dirac occupation $F(\varepsilon(\mathbf{k}))$ is non-vanishing. As such, it strongly depends on the shape of the (occupied) Fermi region is the Brillouin zone (BZ), particularly its co-dimensionality. This intra-band response is the expectation of $\bold v(\bold k+\bold e \bold A(t))$ within the occupied region that is dynamically shifted by the external electromagnetic impulse. When the occupied region is not a ``blob'' in the BZ but a thin Fermi ``tube'' of nontrivial co-dimension, the expectation of $\bold v(\bold k+\bold e \bold A(t))$ can fluctuate wildly with $\bold A(t)$ due to ``destructive interference'' of $\bold v$ at different $\bold A(t)$ shifts, as previously pointed out in Refs.~\cite{mikhailov2007non,lee2015negative,PhysRevB.102.035138}. In Ref.~\cite{mikhailov2007non}, this observation was first used to explain the non-linear response of Graphene due to its vanishing density of states at small chemical potential. This was generalized in~\cite{lee2015negative} to nodal loop structures i.e. where two bands intersect along a loop in the BZ, where a stronger non-linearity from a characteristic non-linear response curve was reported. In Ref.~\cite{PhysRevB.102.035138}, this non-linear response was further shown to be strongly enhanced in a 3D material with non-trivial nodal topology i.e. when band intersections form loops that are topologically linked. 

To more concretely understand why nodal structures with nontrivial linkages have strong non-linear responses, consider small chemical potentials $\mu$ such that the occupied states take the form of nodal ``tubes'' along nodal lines, which are then shifted by the electromagnetic impulse $\int^t\mathbf{E}(t')dt'$ away from the original nodal structure in the BZ. By differentiating Eq.~\ref{J} with respect to the vector potential, we obtain~\cite{PhysRevB.102.035138} for small $\mu$, 
\begin{equation}
\frac{\partial J_i}{\partial A_j}\approx2\sum_{\alpha\in\text{NLs}}\frac{\mu}{\hat{e_j}\cdot\mathbf{v_F}}\frac{\partial^2\varepsilon(\mathbf{k}^\alpha+e\mathbf{A})}{\partial k_i \partial k_j}
\label{dJdA}
\end{equation}
where $\mathbf{k}^\alpha$ labels the trajectories of all the nodal lines $\alpha$ in the BZ and $\hat{e}_j\cdot\mathbf{v}_F$ is the component of the Fermi velocity of the $\alpha$-th NL at momentum $\mathbf{k}^\alpha$ along the applied field. In Eq.~\ref{dJdA}, the differential (optical) response tensor of a nodal line structure is given by a sum over the Hessian of the dispersion at the nodal lines shifted by $\mathbf{A}$ in the BZ. Evidently, we expect the  response to be non-linear whenever $\frac{\partial J_i}{\partial A_j}$ is significantly non-constant, which has to be the case around nodal crossings with vanishing gap. In particular, in the diagonal sector where $i=j$, $\varepsilon(\mathbf{k}^\alpha+e\mathbf{A})$ has to pass through a singularity whenever two nodal loops are linked in the direction $\bold A$, since the locus of $\mathbf{k}^\alpha+e\mathbf{A}$ for one of the loops $\alpha$ must intersect another loop $\alpha'$ as $\bold A$ increases. This leads to the enhancement of diagonal response non-linearity in topologically linked nodal loops, which can be quantified by the extent of HHG it causes.

However, the abovementioned argument for non-linearity enhancement pertains only to the diagonal $i=j$ sector, where the response is probed in the direction where the applied field displaces the nodal loops. It remains an open question if a nodal linkage also possess signature response behaviors in the other directions, in both longitudinal and transverse sectors. In this work, we shall focus on elucidating the signature contributions to the optical response due to a generic nodal linkage, particularly on how non-topological contributions from the dispersion profile competes with those from the topological linkage. While real nodal materials may also experience response contributions from the Berry curvature and inter-band scattering, these additional contributions are very material-dependent, and do not form the focus of this work, which aims to explore the most generic response behavior expected from nodal linkages.

\subsection{Canonical Hopf link Model}
\begin{figure}
\centering
\includegraphics{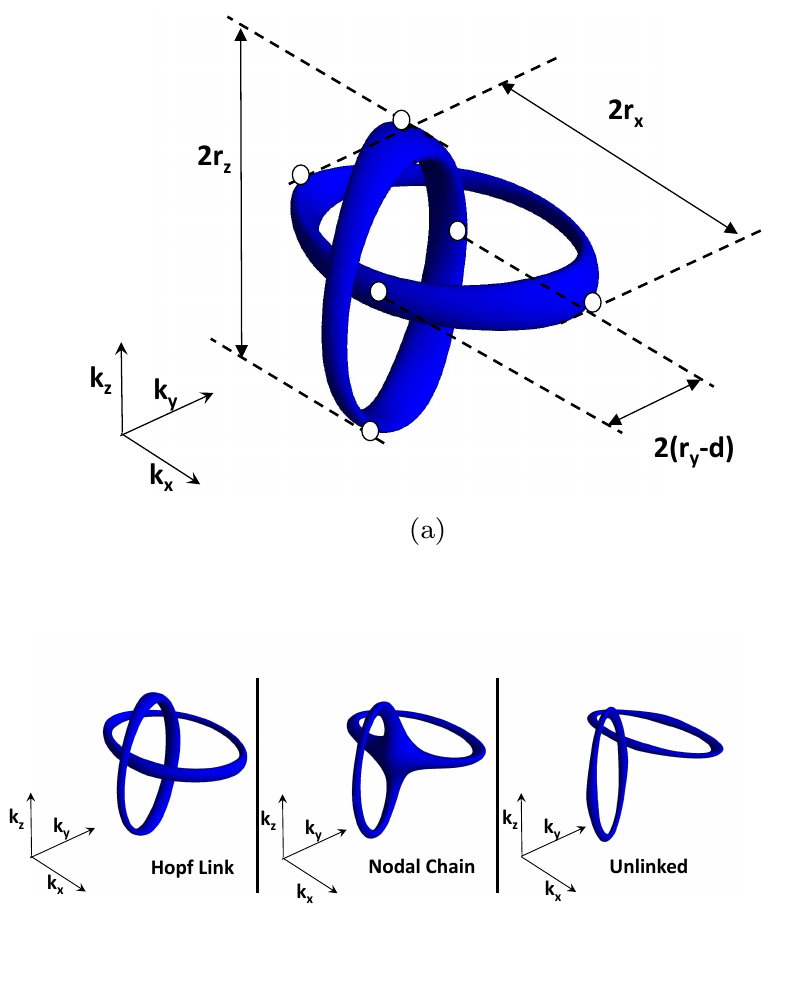}
    \caption{(a) The Fermi surface of the Hopf link in $\mathbf{k}$-space, with the parameters $r_x$, $r_y$, $r_z$ and $d$ as labeled. We assumed $r_x=r_z$. The quantity $2(r_y-d)$ controls the extent of linkage in the Hopf link. (b) The three distinct possibilities of the Hopf structure are: Hopf link ($r_y<d$), nodal chain ($r_y\approx d$) and unlinked nodal loops ($r_y>d$), illustrated here for a non-vanishing $\mu=0.1$.}
    \label{Hopf_shape}
\end{figure}

In this work, we have chosen to focus on the simplest topologically non-trivial nodal linkage, which is the Hopf link - a nodal configuration consisting of two nodal loops that will be linked when they are sufficiently close to each other. By introducing a canonical two-band Hopf link model where the key parameters $r_y$, $r_x$ and $d$ (Fig.~\ref{Hopf_shape}a) are all independently adjustable, we can isolate the various geometric and topological factors that influence its non-linear response. The parameters are:
\begin{itemize}
    \item $r_y$: width of the loops in the $k_y$-direction or longitudinal direction along which the two loops are separated;
    \item $r_x$: width of the loops in the $k_x$ or $k_z$ transverse directions;
		\item $2d$: the displacement between the two loops, centered at $(0,\pm d,0)$. They are linked if $r_y>d$, since the longitudinal separation between them is given by $2(r_y-d)$ (Fig.~\ref{Hopf_shape}).
\end{itemize}

Our two-band nodal Hopf link Hamiltonian takes the form
\begin{eqnarray}H_{\text{hopf}}(\mathbf{k})=h(k_x,k_y,k_z)\sigma_x+g(k_x,k_y,k_z)\sigma_y
\end{eqnarray}
where the functions $h(\mathbf{k})$ and $g(\mathbf{k})$ are obtained by deforming a well-known Hopf link model derived through the Hopf map~\cite{PhysRevB.96.041202,PhysRevB.97.155140,PhysRevLett.118.147003,lee2020imaging}, for which the Hopf link parameters cannot be easily adjusted independently~\footnote{An analogous approach~\cite{lee2017band} with complex singularities also does not allow for independent tuning of parameters.}. The gap $2\epsilon(\mathbf{k})=2\sqrt{g(\mathbf{k})^2+h(\mathbf{k})^2}$ closes along two loops given by the simultaneous solutions of $g(\bold k)=0$ and $h(\bold k)=0$. Explicitly, with the intermediate parameters $\xi=\frac{\cos r_y-1}{\cos r_x-1}\frac{1}{\sin d}-1$ and $\Gamma=\frac{\cos r_x\cos r_y -1}{\cos r_x -1}\frac{1}{\sin d }$, $h(\bold k)$ and $g(\bold k)$ are given by
	\begin{widetext}
	\begin{subequations}	
	\begin{align}
h(k_x,k_y,k_z)&=\sin^2(k_z-k_x)-(\xi\cos(k_x+k_z)+\cot d \cos(k_y)+\cos(k_z-k_x)-\Gamma)^2-\sin^2(k_x+k_z)+\sin^2(k_y),\\
g(k_x,k_y,k_z)&=2\sin(k_z-k_x)(\xi\cos(k_x+k_z)+\cot d \cos(k_y)+\cos(k_z-k_x)-\Gamma)-2\sin(k_x+k_z)\sin(k_y).
	\end{align}
	\end{subequations}
	\end{widetext}
	
At chemical potential $\mu=0$, the two nodal loops lie in the planes $k_x=0$ and $k_z=0$, with a common mirror symmetry axis along the $k_y$ line. The nodal loops each touch this symmetry axis at two points, giving rise to a total of four touching points $\pm(r_y\mp d)$. From Fig.~\ref{Hopf_shape}, $r_y-d$ evidently describes the extent of their linkage.  When $r_y>d$, the two nodal loops are linked together, and we call this resulting arrangement to be a Hopf link. 

In realistic materials, the chemical potential $\mu$ can often be tuned away from the nodal energy, so as to obtain a finite density of states available for transport. For generic linearly dispersive nodal systems, $\mu$ scales with the thickness of the nodal tube. A large $\mu$ tends to result in thick tubes that inevitably intersects, resulting in a nodal chain even when $r_y\neq d$. 

In total, there are three possibilities as illustrated in Fig.~\ref{Hopf_shape}b:
\begin{itemize}
\item Hopf link: The two nodal loops are topologically linked ($r_y>d$ for $\mu\ll 1$),
\item Unlinked nodal loops: The two nodal loops are disjoint and not topologically linked ($r_y<d$ for $\mu\ll 1$),
\item Nodal chain: The two nodal loops touch each other at a point, and are not topologically linked ($r_y\approx d$ for $\mu\ll 1$).
\end{itemize}

Due to the periodicity of its various terms, the Hamiltonian can sometimes admit additional solutions i.e. ``\emph{periodic images}'' within the first BZ. These solutions, analogous to the degenerate valleys in Graphene, occur as two distinct types, as further elaborated in the Appendix. To prevent the confounding ambiguities resulting from the interference of multiple nodal linkages, we shall exclusively study only cases where such periodic images do not exist.

For the rest of this work, we shall set the temperature to a representative value of $10$ K i.e. $\beta:=\frac{eV}{k_BT}=1160$, which can be adjusted to fit the physical temperature of an actual physical scenario by trivially rescaling our canonical model. For definiteness, we choose a chemical potential of $\mu=0.1$ eV, such that the nodal structure consist of thin closed tubes that exhibit distinct crossovers in topology (linked, intersecting, unlinked) as they are shifted due to photon excitations. However, our results shall continue to hold even when typical photon energies are much smaller than $\mu$. We shall study the behavior of the optical response as the three independent parameters $\{d,r_x,r_y\}$ are varied, insofar as the parameters do not admit periodic images, for all three types of cases (linked, unlinked and nodal chain).

\section{Optical Response in various directions}
We next investigate the various diagonal and off-diagonal (Hall) responses of the nodal Hopf link in the principal directions - both longitudinal (along the direction of loop separation $\hat y$) and transverse directions (perpendicular to the direction of loop separation $\hat x$ and $\hat z$). 

Consider a sinusoidal time-varying applied electric field signal
$\mathbf{E}(t)=\mathbf{E_0}\cos\Omega t$, which corresponds to the vector potential $\mathbf{A}(t)=\mathbf{A}\sin\Omega t$, where $\mathbf{A}=\mathbf{E_0}/i\Omega$. In the presence of an oscillatory electric field $\mathbf{E}(t)$, the minimally coupled Fermi-Dirac occupation in Eq.~1 will be $F(\varepsilon(\mathbf{k}-e\mathbf{A}(t)))$ which is equivalent to a translation of the Fermi tube of occupied states in the field direction according to the momentum shift $\mathbf{k}\rightarrow\mathbf{k}-e\mathbf{A}(t).$ This picture motivates the $\mathbf{A}$ dependence of $\mathbf{J}$ rather than $\mathbf{E}$. The ballistic criterion $\Omega\sim100$ THz is independent of the nodal energy since the Fermi surface is translationally invariant. Although there are nine possible response components $J_i(A\hat{j})$, $i,j\in{x,y,z}$, some of them necessarily generically vanish or are not independent due to the symmetries of our Hamiltonian (Eq.~3):
\begin{figure}
\includegraphics[width=.6\linewidth]{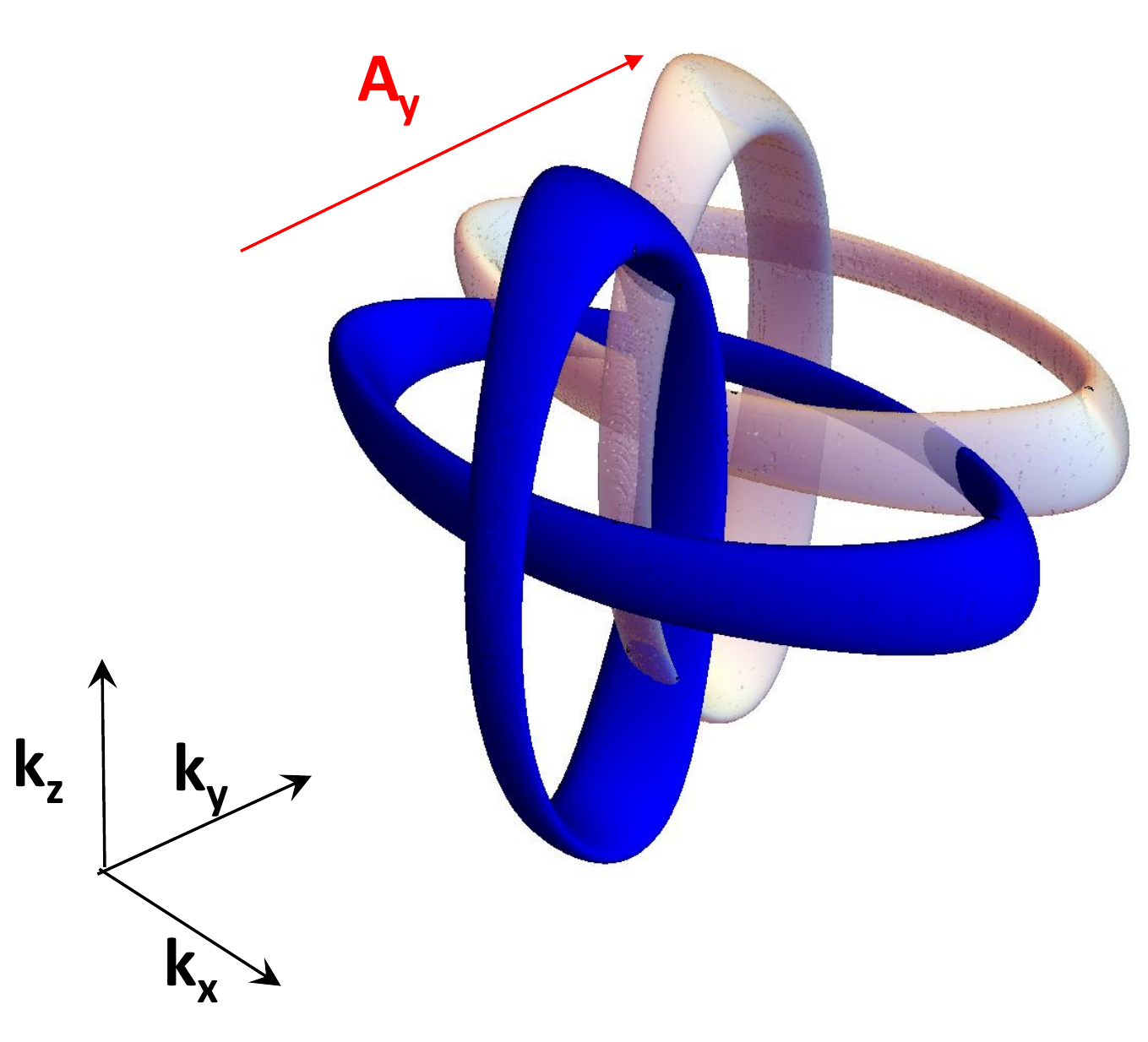}
\caption{Contribution to the semi-classical response (Eq.~1): In the presence of an illustrative applied field along the $k_y$ direction, the Fermi surface is displaced along the $k_y$-axis in $\mathbf{k}$-space in a manner symmetric about the $k_x=0$ and $k_z=0$ planes. This accounts for the vanishing response $J_x$ and $J_z$.}
    \label{vanishing}
\end{figure}
\begin{equation}
    J_x(A\hat{y})=0,\quad J_z(A\hat{y})=0,\tag{5a}
    \end{equation}

    \begin{equation}
    J_x(A~ \hat{x})=J_z(A\hat{z}),\tag{5b}
    \end{equation}

    \begin{equation}
    J_z(A\hat{x})=J_x(A\hat{z}),\tag{5c}
    \end{equation}

    \begin{equation}
    J_y(A\hat{x})=-J_y(A\hat{z}).\tag{5d}
    \end{equation}
\begin{figure*}
\includegraphics{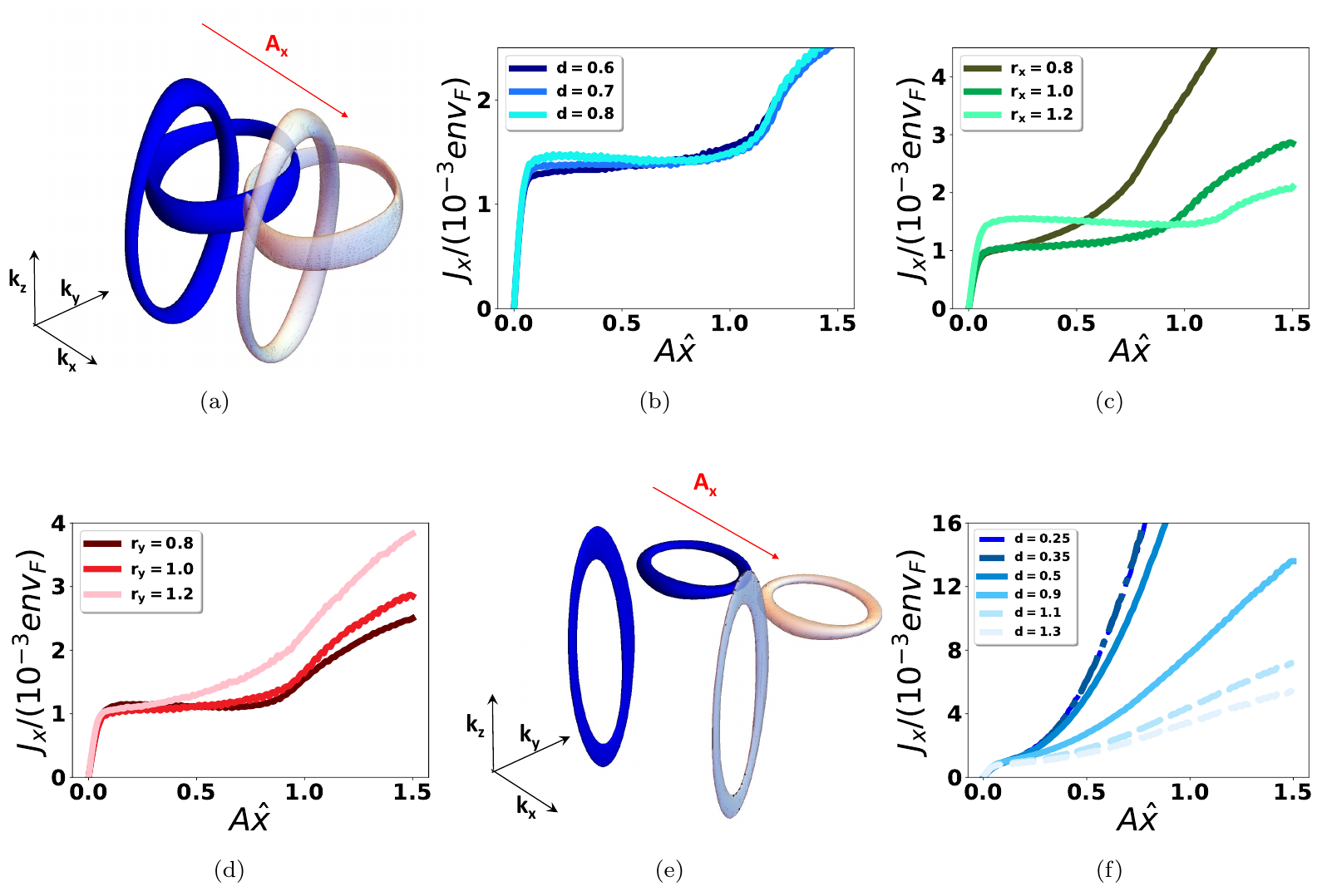}
\caption{(a) In the presence of an applied transverse field $A\hat{x}$, the Fermi surface of the Hopf link is displaced along the transverse $k_x$-axis in $\mathbf{k}$-space, in turn giving rise to a transverse diagonal optical response $J_x(A\hat{x})$. (b) For the Hopf link, the transverse response weakly depends on $d$, for $r_x=0.8$ and $r_y=1.2>d$. (c) Enhancement of transverse response non-linearity for the Hopf link as $r_x$ increases, for $d=0.5<r_y=1.0$. (d) Enhancement of transverse response non-linearity for the Hopf link with $r_y>d=0.5$ and $r_x=1.0$. (e) Displacement of the Fermi regions of the unlinked nodal loops along the transverse $k_x$ axis in $\mathbf{k}$-space, with the singularities from each loop becoming weaker as the loops are further separated. (f) The transverse response shows a stronger dependence on the $d$ when the nodal loops are no longer topologically linked. As the longitudinal loops separation increases, the response currents become smaller. Here, we contrast the responses of the Hopf link (dashed-dot lines, $d=0.25,0.35$), unlinked nodal loops (dashed lines, $d=1.1,1.3$) and the nodal chain (solid lines $d=0.5,0.9$) for $r_y=0.6$, $r_x=0.4$.  
For (b-d), the chosen parameter sets are among those that exhibit the greatest non-linearity in the transverse response. The responses are in units of $env_F$ where $n$ and $v_F$ are the number density and Fermi velocity of the NLSM material.}
\label{Jx}
\end{figure*}

\begin{figure*}
\includegraphics{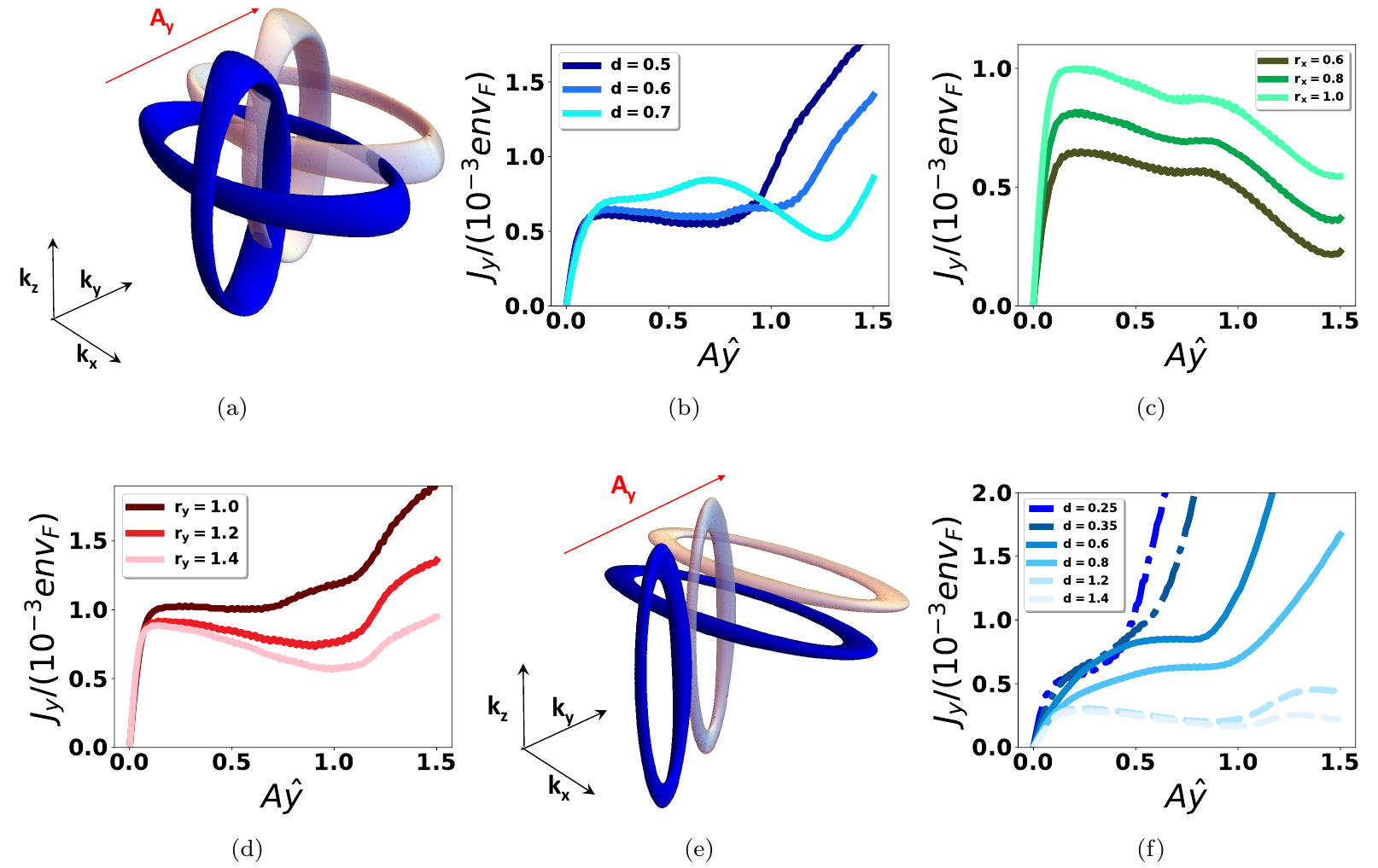}
\caption{(a) In the presence of an applied longitudinal field $A\hat{y}$, the Fermi surface of the Hopf link is displaced along the longitudinal $k_y$-axis in $\mathbf{k}$-space, in turn giving rise to a longitudinal diagonal optical response $J_y(A\hat{y})$. (b) Enhancement of longitudinal response non-linearity by increasing $d$ in the linked regime with $r_y=1.0>d$, $r_x=0.6$. (c) The shape of the longitudinal response for the Hopf link is independent of the transverse radius $r_x$ but increases in magnitude with $r_x$, as plotted for $r_y=1.2>d=0.8$. (d) Enhancement of longitudinal response non-linearity for the Hopf link by increasing $r_y>d=0.6$ at $r_x=1.0$. (e) Displacement of the Fermi regions of the unlinked nodal loops along the longitudinal $k_y$ axis in $\mathbf{k}$-space, with the singularities from each loop becoming weaker as the loops are further separated.  (f) Comparing the longitudinal responses for the Hopf link (dashed-dot lines $d=0.25, 0.35$), unlinked nodal loops (dashed lines $d=1.2, d=1.4$), and the nodal chain (solid lines $d=0.4, d=0.8$) for $r_y=0.6$, $r_x=0.4$. The longitudinal response decreases in magnitude but becomes more non-linear as $d$ increases. This increasing non-linearity is an artifact of the weaker influence of the singularities as the loops are further separated. For (b-d), the chosen parameter sets are among those that exhibit the greatest non-linearity in the transverse response. The responses are in units of $env_F$ where $n$ and $v_F$ are the number density and Fermi velocity of the NLSM material.
}
\label{Jy}
\end{figure*}
Eq.~5a holds since the nodal loops have mirror symmetry about $k_x=0$ and $k_z=0$ planes respectively. As the dispersion velocity $\langle\mathbf{k}|\hat{\mathbf{J}}|\mathbf{k}\rangle$ (Eq.~1) inherits the same symmetry as the Fermi surface - mirror symmetry about the $k_y$ axis,  When the Fermi surface is displaced along the longitudinal $k_y$ axis (Fig.~\ref{vanishing}), the resulting contribution by the Fermi tubes on the left of the $k_y$ axis will exactly cancel that due to the Fermi tubes on the right. Eq.~5b and Eq.~5c follow from the symmetrical roles played by transverse momenta $k_z$ and $k_z$. Eq.~5d is true because of the loops are related to each other by interchanging $k_x\leftrightarrow-k_z$.  In all, there are only 4 unique currents responses in the following response matrix 
\begin{equation}	
J_i(A\hat{j})=\begin{pmatrix}J_x(A\hat{x})&J_y(A\hat{x})&J_z(A\hat{x})\\0&J_y(A\hat{y})&0\\J_z(A\hat{x})&-J_y(A\hat{x})&J_x(A\hat{x}).\\
\end{pmatrix}
\end{equation}	
Since $A$ scales with $E_0$ for our sinusoidal electric field, Eq. 5 is similar to a conductivity tensor. As we shall show, the diagonal responses are much more affected by the topological linkage of the nodal loops, compared to the non-diagonal response, i.e. Hall responses. 

\subsection{Transverse Diagonal Responses $J_x(A\hat{x})$, $J_z(A\hat{z})$}
According to Eq. 1, we generically expect a non-linear semiclassical response since its integral is a highly non-linear function. This is especially the case for a Hopf link, where each nodal loop juts out of the plane of the other nodal loop. As such, it acts as a source of dispersion velocity $\langle\mathbf{k}|\mathbf{\hat{J}}|\mathbf{k}\rangle$, which `destructively interferes' with those due to the nodal loop that encloses this nodal tube~\cite{PhysRevB.102.035138}. When the Fermi surface of the Hopf link starts to translate by the impulse from an applied field (Fig.~\ref{Jx}a), the response, computed from Eq.~1, initially increases sharply followed by a much slower non-monotonic change until the applied field is sufficiently large that the Fermi surface begins to leave the region of influence exerted by the singularity. 

Beyond that, the contribution of $\langle\mathbf{k}|\mathbf{\hat{J}}|\mathbf{k}\rangle$ continues to add up, giving a subsequent monotonic response. The extent of non-linearity afforded by each singularity is described by Eq.~2, where the gradient is a sum of the second derivatives of the dispersion. The transverse (perpendicular to the direction of the loop separation) diagonal optical response of the Hopf link was found to observe the following general trends (we plot the most non-linear responses in Figs.~\ref{Jx}b-d):
\begin{enumerate}
    \item The response demonstrates a weak dependence on the longitudinal separation $d$ when the nodal loops are topologically linked,
     but not true otherwise, as illustrated in Figs.~\ref{Jx}b and f respectively. 
    \item The extent of response non-linearity is enhanced with larger transverse radius $r_x$ or smaller longitudinal radius $r_y$ i.e. larger loop aspect ratio $r_x/r_y$, as illustrated in Figs.~\ref{Jx}c,d. Correspondingly, the response currents generally decrease with increasing non-linearity.
\end{enumerate}
We can physically understand these trends. Since the relative separation $d$ of the centers of the loops merely changes the relative longitudinal $k_y$ position of the loops, intuitively it should not play a significant role in influencing the transverse response resulting from the transverse translation of the Fermi surface. Indeed, the transverse diagonal response is almost independent of $d$ (Fig.~\ref{Jx}b) in the Hopf link. Yet, this independence from $d$ no longer holds when the loops are no longer topologically linked (Figs.~\ref{Jx}e,f) because further separated loops i.e. sources of singularities correspond to more uniform velocity fields, leading to smaller responses currents. This is unlike the linked cases, where the nodal linkage guarantees the proximity to nodal singularities and topologically ``protects'' the transverse response (previous literature i.e. Ref.~\cite{PhysRevB.102.035138} only reported the topological enhancement of the longitudinal response).

On the other hand, increasing the transverse radius $r_x$ or decreasing the longitudinal radius $r_y$ stretches the aspect ratio of each loop and hence the boundary for ``interference'' (defined by the locus of points where the competition of opposing dispersion velocity vectors terminate) along the transverse $k_x$ direction, giving more room for ``destructive interference'' as the Fermi surface is displaced along the transverse $k_x$ direction. This in turn gives a  greater range of $A\hat{x}$ where the response undergoes a non-monotonic change.

\subsection{Longitudinal Diagonal Response $J_y(A\hat y)$}

Unlike in the transverse diagonal response, the longitudinal separations of the loops $2(r_y-d)$ now greatly influences the longitudinal (along the direction of the loop separation) responses of the Hopf link. Like before, these nodal tubes are sources of `destructive interference' for the dispersion velocity $\langle\mathbf{k}|\mathbf{\hat{J}}|\mathbf{k}\rangle$. The semiclassical response is again computed from Eq. 1 as the Fermi surface of the Hopf link displaces along the longitudinal $k_y$ direction (Fig.~\ref{Jy}a). The general trends for the longitudinal diagonal optical response of the Hopf link are:
\begin{enumerate}
    \item The extent of response non-linearity is enhanced with larger longitudinal loop separation $d$, as illustrated in Fig.~\ref{Jy}b.    
		\item Increasing the longitudinal radius $r_y$ also enhances the extent of response non-linearity, as illustrated in Fig.~\ref{Jy}d.
    \item However, the transverse radius $r_x$ does not change the shape of the response curves, but the magnitude of the response does increase with $r_x$, as illustrated in Fig.~\ref{Jy}c.

\end{enumerate}

To physically explain these trends, note that as $d$ increases with $r_y$ fixed, the loop separation $2(r_y-d)$ decreases, enhancing the extent of non-linearity due to the closer proximity of one loop with the band singularity of the other. While the non-linearity is still present, as shown in Fig.~\ref{Jy}f, when the nodal loops are no longer topologically linked, this is an artifact from the weaker dispersion field when the loops are separated farther apart. Increasing $r_y$ at fixed $d$ also enhances non-linear response by providing a greater range of values of $A~\hat{y}$ for significant destructive interference to occur. 

On the other hand, $r_x$ only stretches the nodal loop in the transverse $k_x$ direction, which plays no role in the `destructive interference' in the longitudinal direction. But yet, $r_x$ increases the circumferential length of the nodal loop, which gives a greater contribution to $\mathbf{J}$ in Eq.~1 and hence a greater response magnitude. 
 
 \section{Global aspects of optical response anisotropy}
\begin{figure}
    \centering
 \includegraphics{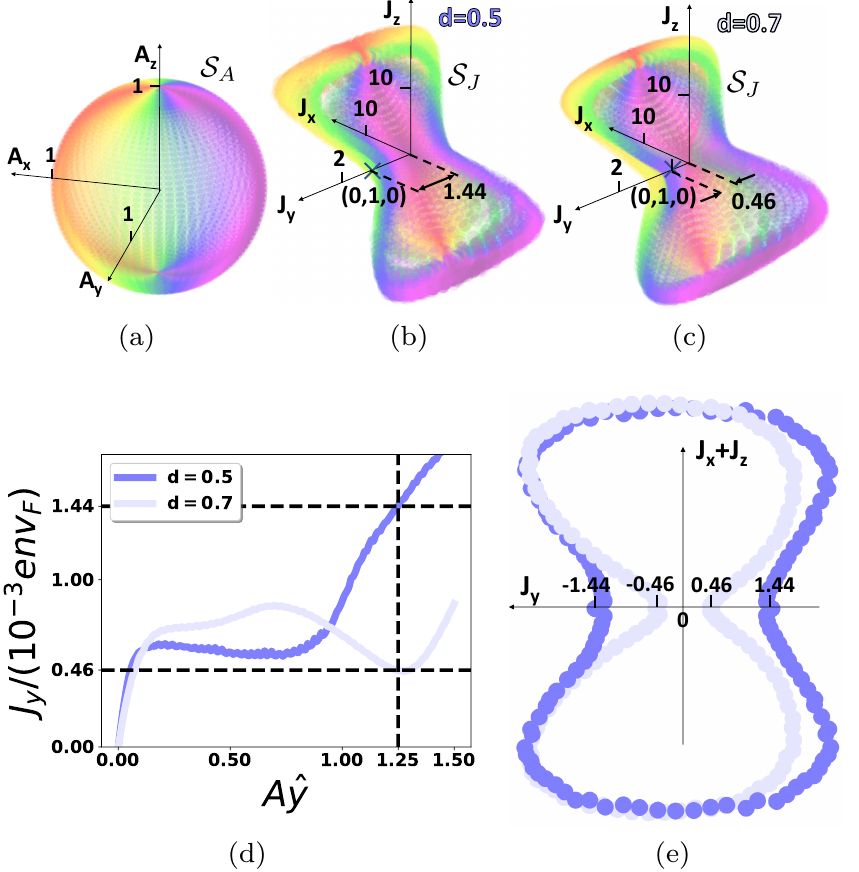}
    \caption{
    (a) To draw the response surfaces, we represent an external field of constant magnitude $A_0$ but arbitrary direction as a sphere $\mathcal{S}_A$ in $A$-space. The sphere is colored by the azimuthal angle $\phi$, and is significantly distorted by anisotropy when bijectively mapped onto the response surfaces $\mathcal{S}_J$ in 3D $J$-space. (b-c) Illustrative examples of the distortion for our model (Eq.~3) with $r_x=0.6,r_y=1.0,\mu=0.1,A_0=1.25$, for $d=0.5$ and $0.7$ respectively ($\mathbf{J}$ in units of $10^{-3}env_F$). (d) Origin of the thicker ``waist'' of (b) vs. (c) in terms of $J_y(A_0 \hat y)$ response. At $A_0=1.25$, the $d=0.5$ case (b) has a response current almost thrice as large. 
     (e) This difference in $J_y(A_0 \hat y)$ corresponds to a large difference in the thickness of the concave ``waist'' of the response surface, as more clearly illustrated by comparing cross sections of (b,c) at $\theta=\frac{\pi}{2}$. }
\label{A0}
\end{figure}

\begin{figure*}
\includegraphics{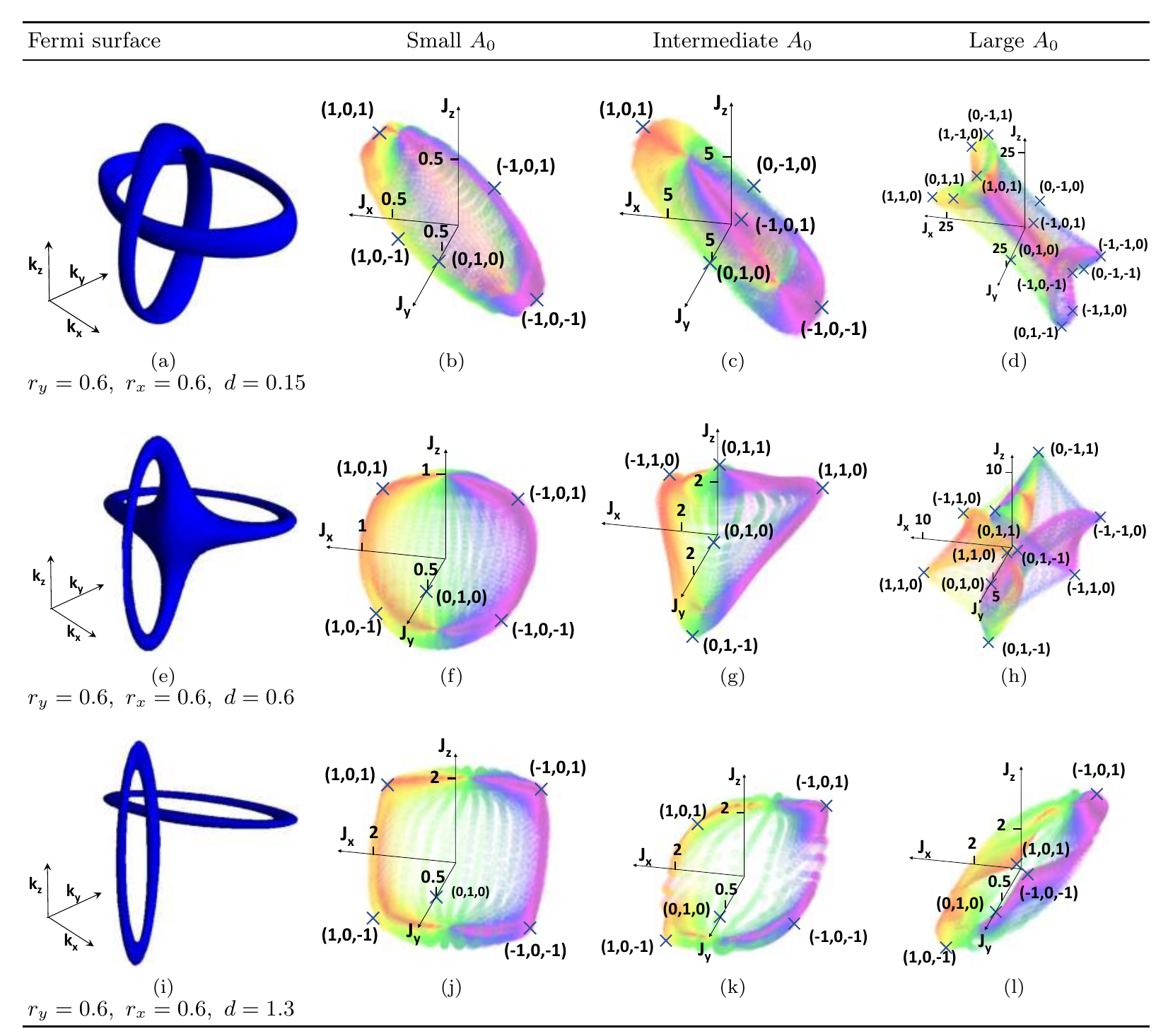}
 \caption{
 For the three possible nodal configurations (linked, nodal chain, unlinked) of our double loop Hamiltonian, shown here are their corresponding surfaces $\mathcal{S}_J$ for constant chemical potential $\mu=0.1$ under small, medium and large impulses $A_0=0.1,0.75,1.75$ for the linked and chain case, while $A_0=0.1,0.5,1.0$ for the unlinked case. The qualitative evolution of the response surfaces with increasing $A_0$ differ for the different topological nodal configurations, with the large-$A_0$ response surface elongated towards the $J_z\pm J_x$ direction in the topologically linked/unlinked cases.
 (a-d) The response surfaces for the Hopf link $r_y=0.6>d=0.15$, $r_x=0.6$ exhibit concavity along directions $\pm(0,1,1)$, $(1,0,\pm1)$ due to the locally depressed responses in these directions. (e-h) The response surfaces for the nodal chain $r_y=0.6=d$, $r_x=0.6$ do not obey the same symmetries as those in the Hopf link since the Fermi surface of the nodal chain breaks reflection symmetry. In particular, for intermediate $A_0$ values, the surface looks like a saddle with asymmetric responses, e.g. $\hat{A}=(0,1,1)$ has a greater response than $\hat{A}=(0,-1,-1)$ at $A_0=0.75$. (i-l) The response surfaces for unlinked nodal rings $r_y=0.6>d=0.15$, $r_x=0.6$ do respect these symmetries too but possess topologies distinct to the topologically linked case. The units of the response currents $\mathbf{J}$ are $10^{-3}env_F$, and parameters are chosen such that no periodic images are present. }
 
\label{A1}
\end{figure*}

\begin{figure*}
\includegraphics{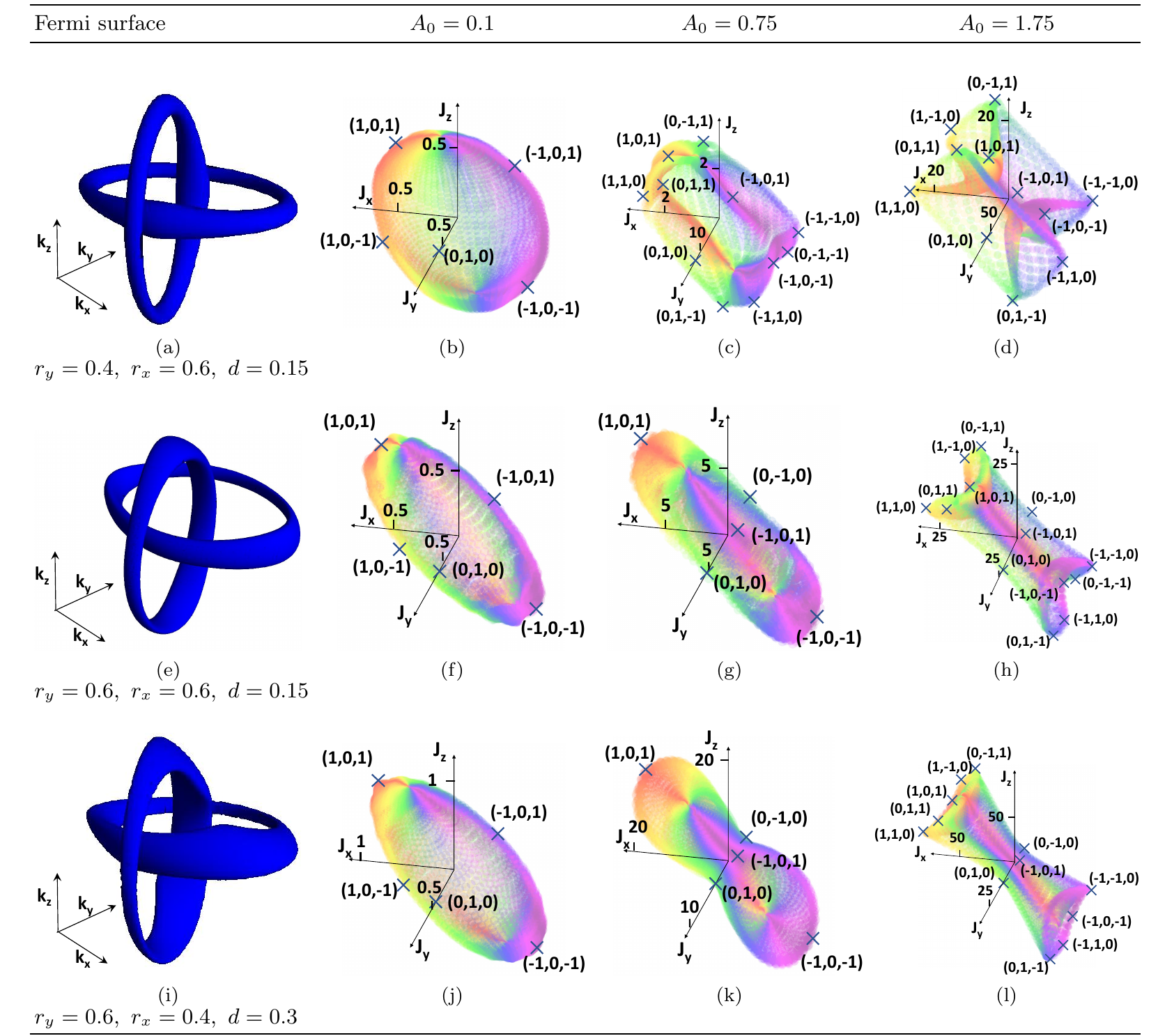}
\caption{The dependence of the surface morphologies on the aspect ratio parameters $r_x,r_y$ of the nodal Hopf link at constant $\mu=0.1$. Again, their response surfaces $\mathcal{S}_J$ are plotted for small, medium and large impulses $A_0=0.1,0.75$ and $1.75$, with current $\mathbf{J}$ are in units of $10^{-3}env_F$. (a-d) The response surfaces for the Hopf link $r_y=0.4>d=0.15$, $r_x=0.6$. (e-h) The same response surfaces as in Fig.\ref{A1}a-d. (i-l) The response surfaces for the Hopf link $r_y=0.6>d=0.15$, $r_x=0.4$ share several similarities with (e-h), but with a more depressed longitudinal response at $A_0=0.75$ and $A_0=1.75$ and a less depressed response along $\hat{A}=\pm(1,0,1)$ at $A_0=1.75$. Overall, the response surfaces differ for different geometric parameters, but do not change as drastically as when the nodal topology changes (Fig.~6).}
\label{A2}
\end{figure*}

Having discussed the non-linearity of the response tensor along the principal directions, we now present how the anisotropic response behaves as a whole. Due to non-linearity i.e. $\bold J(\bold A_1+\bold A_2)\neq \bold J(\bold A_1)+\bold J(\bold A_2)$, the responses in directions away from the previously studied principal directions $\bold{\hat{A}}=(1,0,0)$, $(0,1,0)$ and $(0,0,1)$ may behave unexpectedly. In the following, we shall represent the response across all directions in the form of constant $|\bold A|=A_0$ level surfaces in $\bold J$-space (recall that $\bold A(t)=\int_{-\infty}^{t}\mathbf{E}(t')dt'$).  We parametrize this surface $\mathcal{S}_A$ with spherical angular coordinates as follows:
\begin{equation}
\mathcal{S}_A=\{A_0\boldsymbol{\hat{A}}=A_0(\cos\theta\cos\phi,\cos\theta\sin\phi,\sin\theta)\}\tag{8}
\end{equation}
where $0\leq\theta\leq\pi,~0\leq\phi\leq 2\pi$. The anisotropy in the optical response of our canonical double nodal loop model (Eqs.~3, 4a,b) due to $\boldsymbol{\alpha}=A_0\boldsymbol{\hat{A}}$ is described by the anisotropy of the smooth $\mathcal{S}_J$ surface embedded in the $J$-space. We represent the azimuthal angle $\phi$ with a colormap defined by $\mathcal{S}_A$ in Fig.~\ref{A0}a, which will be significantly distorted by anisotropy when bijectively mapped onto $\mathcal{S}_J$ (Figs.~\ref{A1} and \ref{A2}). 

Insight into the shapes of of the constant $A_0$ response surfaces, as well as their significance with regards to nodal topology and geometry, can be obtained by computing the response in a few high-symmetry directions of $\boldsymbol{\hat{A}}$, such as $\boldsymbol{\hat{A}}=\pm(0,1,0),\pm(1,0,1),\pm(1,0,-1)$ (for notational simplicity, we shall henceforth drop the normalization factor). From Eq. 5a, we know that for a longitudinal external field $\boldsymbol{\hat{A}}=\pm(0,1,0)$, there is vanishing transverse response, i.e. $J_x=J_z=0$ for such $\boldsymbol{\hat A}$. Hence the corresponding point on the $\mathcal{S}_J$ surface must be along the $J_y$ axis, with $|J_y|$ being the width of the surface along the $J_y$-axis. This is also the longitudinal diagonal current discussed in Sect. III. Also, the response surface point corresponding to $\boldsymbol{\hat{A}}=\pm(1,0,1)$ must lie in the $J_x-J_z$ plane, since $J_y=0$ from Eq. 5d. And by Eq. 5b, there is an equal contribution of $J_x$ and $J_z$ from $\boldsymbol{\hat{A}}=\pm(1,0,1)$, so all corresponding points from this orientation of $\boldsymbol{\hat A}$ must lie on the 45 degree diagonals along the $J_x$-$J_z$ axes, as evident in Figs.~\ref{A1} and \ref{A2}. Incidentally, points corresponding to $\boldsymbol{\hat A}=\pm(1,0,-1)$ also lie in the $J_x$-$J_z$ plane despite not being directly protected by Eqs.~5a and 5d, a fine illustration of the non-linearity of $\bold J(\bold A)$. Other high symmetry directions which we shall use in the following discussion include $\boldsymbol{\hat{A}}=~\pm(1,1,0),~\pm(0,1,1),~\pm(1,-1,0),~\pm(0,1,-1)$.
 
Before starting the detailed analysis of the response surface, we consider a quick example. In Sect. III, we highlighted that a greater longitudinal separation $d$ between two topologically linked nodal loops results in a more non-linear response (Fig.~\ref{Jy}b). This is indeed seen in our corresponding $\mathcal{S}_J$ surfaces. At an illustrative $A_0=1.25$, this non-linear response is approximately $3$ times larger i.e. less non-linear for $d=0.5$ vs $d=0.7$ (Fig.~\ref{A0}d), which corresponds to a proportionally smaller width of the ``waist'' of the response surface in the $\hat y$ direction (Figs.~\ref{A0}b,c). This is more evident when viewed in the cross-sectional plane $\hat{A}=(1,0,1)$ with only $\theta=\frac{\pi}{2}$ points plotted (Fig.~\ref{A0}e). That the general shapes of the $d=0.5$ and $d=0.7$ surfaces appear rather similar (Figs.~\ref{A0}b,c) can be understood from the weak dependence on $d$ for the transverse diagonal responses, as discussed in Sect. III. Clearly, the results from Sect. III alone are inadequate in working out the entire response surface and thus the response surfaces provide a bigger picture and as we will see, shed light on nodal topology and geometry.

\subsection{Response surfaces and nodal topology}
It is already known that~\cite{PhysRevB.102.035138} nodal linkages can enhance the non-linearity of the response, at least in the longitudinal diagonal direction $J_y(A_0\hat{y})$ (Fig.~4). Further, significant non-linearity is also present in the transverse diagonal directions $J_x(A_0\hat{x})$ and $J_z(A_0 \hat{z})$, although not necessarily enhanced by the topological linkage. Hence we shall expect such non-linearity to be manifested in the constant $A_0$ response current surface too. 

Indeed, as shown in Fig.~6, typical linked, touching and unlinked nodal loops exhibit significantly different evolutions of the constant $A_0$ response surfaces as $A_0$ is increased. At small $A_0$, all three cases have ellipsoidal-shaped responses surfaces, testimony to anisotropy of the nodal system, even in the linear (small $A_0$) limit. At very large $A_0$ i.e. $A_0=1.75$ where the Fermi regions have been displaced far from their original positions, the response surfaces are all very anisotropic and large, since minimal cancellation of the velocity field $d\varepsilon/d\bold k$ occurs. Their exact shapes depend on the details of the energy dispersion away from the loops, and are non-universal though decidedly anisotropic. What is most interesting is the intermediate $A_0\approx 0.75$ regime, which for the linked case is around when the Fermi region of one loop crosses the singularity from the other loop. For the linked case, the significant non-linearity of the response around intermediate values of $A$ (see Figs.~3 and 4) suppresses the response current, particularly in the longitudinal directions with ``untwisted colors'' in Fig.~7. As such, this leads to a somewhat ``squeezed'' appearance of the response surface compared to that of the nodal chain or unlinked cases, where the response surface looks comparatively ``puffed up'' during the $A_0$ evolution. Compared to the other cases, the unlinked case shows the least variation in response surface shape during the evolution due to the least amount of cancellation of $d\varepsilon/d\bold k$ during the evolution.

A more detailed characterization of the shape of the response surfaces can be performed by analyzing the high symmetry directions. For instance, for the Hopf linked case, the constant $A_0$ surfaces look similar to ellipsoids (Fig.~\ref{A1}b,~\ref{A2}b,f,j) with the longest axis oriented along the 45 degree diagonal between $J_x>0$ and $J_z>0$ axes, corresponding to the direction $\bold J\propto \pm(1,0,1)$. The three `principal axes' of the response surfaces are thus marked out by the three pair of points $\pm(0,1,0)$, $\pm(1,0,1)$ and $\pm(1,0-1)$, as discussed. This is seen in Figs.~\ref{Appendix: ry06rx06response}m-p where the magnitude for the responses along $\pm(1,0,-1)$ are significantly smaller than along $\pm(1,0,1)$ and $\pm(0,1,0)$. This accounts for its oblate appearance.

As we increase $A_0$ for the linked case, the two faces of the constant $A_0$ surface in Fig.~\ref{A1}c which are characterized by $\pm(1,0,-1)$, exhibit concavity, reminiscent of the shape of a red blood cell. The concavity along the directions $\pm(1,0,-1)$ can be better seen in Fig.~\ref{Appendix: diff}b, where the same response surface is viewed from a lateral direction. This characteristic suppressed response is similar to that in Fig.~\ref{A0}d where it was a consequence from the non-monotonicity of the response. In these cases, the responses along $\pm(1,0,-1)$ increase significantly slower than in other directions (Figs.~\ref{Appendix: ry06rx06response}m-p). As $A_0$ continues to increase, the $\hat A$ directions characterized by $\pm(1,0,1)$ and $\pm(0,1,0)$ also exhibit concavity. Again, this is because the responses along $\pm(1,1,0)$, $\pm(0,1-1)$, $\pm(0,1,1)$, $\pm(1,-1,0)$ (which correspond to the eight corners of the surface in Fig.~\ref{A1}d) grow much quicker than along the principal directions, consistent with the individual response curves in Figs.~\ref{Appendix: ry06rx06response}a-l. This surface (Fig.~\ref{A1}d) also show symmetry consistent with the symmetries in the responses (Figs.~\ref{Appendix: ry06rx06response}a-l). For instance, the constant $A_0$ surface has mirror symmetry about the $J_y=0$ plane. Due to the non-negligible thickness of the nodal loops, certain symmetries of the $\mu=0$ nodal system may be broken. This symmetry breaking is particularly pronounced in the nodal chain case, where the two nodal rings intersect with relatively weak dispersion. Generically, the finitely  thick nodal tube can break reflection symmetry in the direction of impulse, i.e. translating the Fermi surface in the $+A_0\hat{y}$ direction results in a different response compared to doing so in the $-A_0\hat{y}$ direction. For small $A_0$, the surfaces are imperfectly rounded blobs (Fig.~\ref{A1}f) and evolve to a saddle shape (Fig.~\ref{A1}g) as $A_0$ grows. This saddle-like shape is more obvious when viewed in a different orientation illustrated in Fig.~\ref{Appendix: diff}c. It can be characterized by 6 out of the 14 points in total: $(0,1,1)$, $(1,1,0)$, $(0,\pm1,0)$, $(0,1,-1)$ and $(-1,1,0)$. Again, we observe asymmetry in the sense that the response grow more than proportionately in the directions $(0,1,1)$, $(1,1,0)$, $(0,1,-1)$ and $(-1,1,0)$ (the negative pair does not grow as fast). As $A_0$ grow, so do the responses corresponding to the remaining 8 points, giving to a surface like in Fig.~\ref{A1}h, with 6 concave sides.

For the unlinked nodal loops, the surface for small $A_0$ is again imperfectly rounded due to linear anisotropy. For intermediate $A_0$, the surface expands into a lemon-like shape due to the relatively small response non-linearity, and can be characterized by the surface directions $\pm(1,1,0),\pm(0,1,1)$ and $\pm(1,-1,0)$. As $A_0$ grow, the responses in the $\pm(1,0,1)$ directions do not grow as fast, and thus the face characterized by this pair of points exhibit concavity. For large $A_0$, the response surface elongates towards the $J_z-J_x$ direction which differs from the linked case where it elongates towards the $J_z+J_x$ direction. This conclusively relates the response surfaces with the topological linkage of the nodal loops.

\subsection{Response surfaces and nodal geometry}

Since the non-linear response current does not correspond to any topologically quantized value, we expect it to be affected by deformations of the nodal structure too. This should apply to both linked and unlinked cases, even if the linked case is more likely to possess strongly non-linear response. Earlier, we have considered circular nodal loops with $r_x=r_y$. How will the surfaces in Fig.~\ref{A1} change as we vary $r_y$ and $r_x$? Varying $r_y$ and $r_x$ add one more layer of complexity in our analysis of these surfaces. To study that, we shall start from the parameters in Fig.~\ref{A1}(a-c) and change $r_y$ and $r_x$ separately, as presented in Fig.~7. 

Generally, the surfaces along the same column of Fig.~\ref{A2} will have relatively similar morphologies since they are only slightly distorted from each other and correspond to the same $A_0$ impulse. In particular, the special symmetries in Figs.~\ref{A1}b-d are no longer satisfied as the relative responses along the various directions now behave quite differently. This occurs because a chemical potential of $\mu=0.1$ result in nodal loops with significantly non uniform and asymmetric thickness (Fig.~\ref{A2}i). This further enhances the anisotropy of the response surfaces, particularly at $A_0=0.75$. But at high fields $A_0=1.75$, the surface morphologies of Figs.~\ref{A2}d,h share similar morphological features - suppressed response at $\pm(1,0,1)$ and $\pm(1,0,-1)$. Again, this anisotropic response is due to the varying rates of growth for the responses along the individual directions. For instance, $\pm(1,0,1)$ does not grow as fast as say $\pm(1,1,0)$ in Fig.~\ref{A2}h, but grow at comparable rates in Fig.~\ref{A2}l.

\section{Conclusion}
In this work, we have systematically studied in detail the anisotropic and non-linear optical response of two nodal loops that are linked, unlinked or touching (nodal chain). This system, as parametrized by our canonical two nodal loop model, represents the simplest abstraction of simultaneously occurring nodal loops (linked or unlinked) in nodal materials. First, we studied the effects of nodal geometry and topology individually along various axis directions. Next, we presented constant $A_0$ response surfaces to highlight the anisotropy of the response, and how that global picture can shed light on the overall configuration of the nodal structure. Our findings generalizes existing results on the enhancement of optical response non-linearity by nodal linkages~\cite{PhysRevB.102.035138} to various transverse, Hall and diagonal sectors, and introduces a geometric picture of response non-linearity and anisotropy that will be invaluable in analyzing generic nodal material responses, as well as the engineering of higher-harmonic generation materials for applications like Terahertz radiation generation.

\section{Acknowledgements}
TT is supported by the NSS scholarship by the Agency for Science, Technology and Research (A*STAR), Singapore.
This work is supported by the Singapore Ministry of Education (MOE) Tier 1 start-up grant (R-144-000-435-133).

\bibliography{biblio}

\newpage
\appendix

\section{Periodic images of the nodal Hamiltonian}
\begin{figure*}
\includegraphics{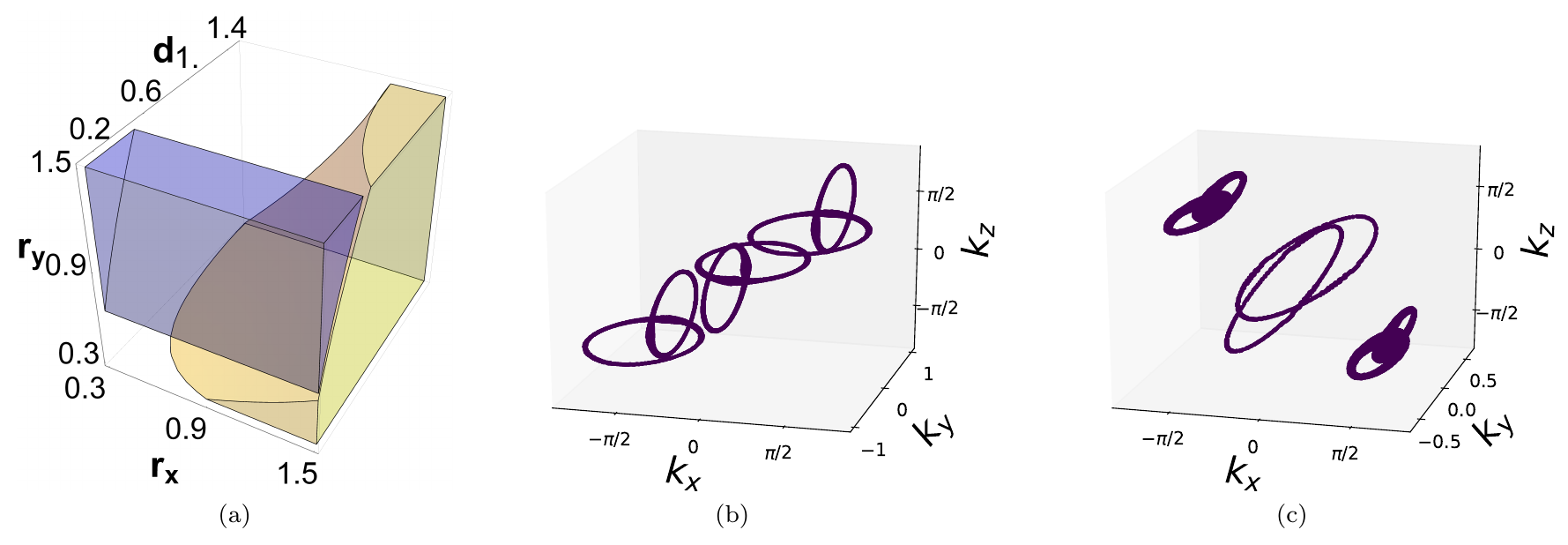}
    \caption{(a) The regions in parameter space for the single Hopf structure (uncolored), Type I periodic images (yellow) and Type II periodic images (purple).  (b-c) For non-vanishing $\mu$, we illustrate the two types of periodic images which come in pairs in the first BZ. (b) The nodal loops of Type I periodic images lie in the planes $(k_x=+\pi/2,~k_z=+\pi/2)$ and $(k_x=-\pi/2,~k_z=-\pi/2)$, as illustrated for the case $r_y=0.6>d=0.4$,  $r_x=1.0$. 
   (c) The nodal loops of Type II periodic images lie in the planes $(k_x=-\pi/2,k_z=+\pi/2)$ and $(k_x=+\pi/2,~k_z=-\pi/2)$, as illustrated for the case $r_y=1.0>d=0.2$, $r_x=0.8$.
}
    \label{periodic}
\end{figure*}

Periodic images occur when there are more than one branch of solution (usually three in total) when we solve $h=0$ and $g=0$ simultaneously in the Hamiltonian $H_\text{hopf}(\bold k)=h(\bold k)\sigma_x +g(\bold k)\sigma_y$, as defined in Eq.~4 of the main text. We can deduce their positions from the $k_z$ planes the periodic images lie in, as well as, the $k_x$ lines these periodic images are symmetrical with respect to. There are two types of periodic images, both which occur in pairs. We distinguish them based on their positions:
\begin{itemize}
    \item Type I: the pair of images respectively have their loops lie in the planes $k_x=+\pi/2$, $k_z=+\pi/2$ and $k_x=-\pi/2$, $k_z=-\pi/2$ (Fig.~\ref{periodic}b);
    \item Type II: the pair of images respectively have their loops lie in the planes $k_x=+\pi/2$,  $k_z=-\pi/2$ and $k_x=-\pi/2$, $k_z=\pi/2$ (Fig.~\ref{periodic}c).
\end{itemize}
Only the single Hopf regime (uncolored in Fig.~\ref{periodic}a) is studied in detail in this work, since it provides the most conclusive results on how the nodal shape and topological linkage affect transport properties.

\begin{figure*}
\centering
 \includegraphics{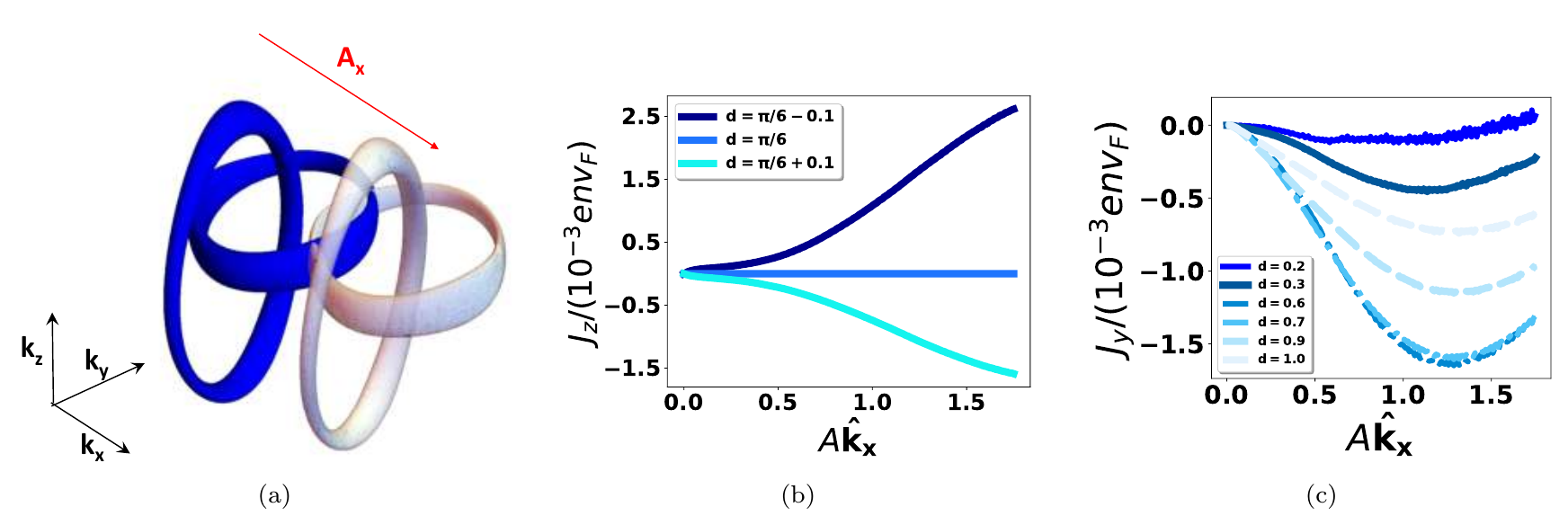}
\caption{(a) In the presence of an applied transverse field $A\hat{x}$, the Fermi surface of the Hopf link is displaced along the transverse $k_x$ axis in $\mathbf{k}$ space. On top of the diagonal responses, we obtain additional Hall responses, which exhibit trends that are not related to the topological linkage of the nodal rings. (b) When the intermediate parameter $\xi$ in the Hamiltonian (Eq. 3) is set to 1, the transverse Hall response will vanish for all values of impulse $A(t)=e\int_{-\infty}^tE(t')dt'$. For the case of circular nodal loops $r_x=r_y$, the corresponding critical $d$ value is $d_{\text{crit}}=\pi/6$. In the plot, we chose $r_x=r_y=0.6$ and $\mu=0.1$ such that the nodal tubes are thick enough to give nodal chains even when $d=\pi/6<r_y$. Yet, even as the nodal loops touch each other, the transverse Hall response still vanishes when $d=d_{\text{crit}}$. This demonstrates the topological linkage of the nodal loops are not important for this response. (c) Comparing the longitudinal transverse Hall responses for the Hopf link (solid lines $d=0.2, 0.3$), unlinked nodal loops (dashed lines $d=0.9, d=1.0$), and the nodal chain (dashed-dot lines $d=0.4, d=0.8$) for $r_y=0.6=r_x$. The response increases in magnitude with $d$ and exhibits the largest magnitude when the nodal loops touch each other. For the Hopf link, this response is extremely weak at small fields and increases as we increase the field strength $A$. Here, there is no distinct difference between the topologically trivial case and the topologically linked case, but rather a smooth continuous change with $d$. (b-c) The responses are in units of $env_F$ where $n$ and $v_F$ are the electronic number density and Fermi velocity of the NLSM material.
}
\label{Hall}
\end{figure*}

 Trivially, $g=0$ in both types of constraints, and we only need to solve for $h=0$ to obtain the explicit form of the periodic images. For Type I images, we have
\begin{equation}
\alpha\sin k_y=\pm(\beta\cos k_y+\gamma)\tag{A1}
\end{equation}
where $\alpha=(\cos r_x-1)$, $\beta=-(\cos r_x-1)\cot d $ and $\gamma=2+\cosec d (-2+\cos r_y)+\cos r_x(\cos r_y\cosec d -2)$. Eq. A1 can be solved as follows:
\begin{align}
& \sqrt{\alpha^2+\beta^2}\sin(k_y\mp\delta)=\pm\gamma\implies \nonumber\\& k_y=\sin^{-1}\bigg(\pm\frac{\gamma}{\sqrt{\alpha^2+\beta^2}}\bigg)\pm\tan^{-1}\frac{\beta}{\alpha}\tag{A2}
\end{align}
where $\frac{\gamma}{\sqrt{\alpha^2+\beta^2}}=\frac{\gamma\sin d}{\cos r_x-1}$, $\frac{\beta}{\alpha}=-\cot d$. 
Type I periodic images arise when there exists real solutions for Eq. A2. This implicitly requires the argument in the inverse sine in Eq. A2 to have magnitude less than 1. In another words, the criteria for type I periodic images is given as the following inequality:
\begin{equation}
\frac{-2+\cos r_y (1+\cos r_x )}{\cos r_x-1}-2\sin d\leq1.\tag{A3}
\end{equation}
Type II images arise from another solution branch described by a quadratic equation in $\cos k_y$, which gives the solution 
\begin{align}
&\cos k_y=\frac{1}{2}\sin^2d\bigg( (4+2\cos r_y)\cot d\cosec d \pm\nonumber\\& \sqrt{-12+4\cot^2d-16\cos r_y\cosec d-4\cos^2 r_y\cosec^2d}\bigg)\tag{A4}
\end{align}
The corresponding regime for Type II image is set by the above discriminant (argument in the square-root term) being strictly positive, i.e.
\begin{equation}
-12+4\cot^2d-16\cos r_y\cosec d-4\cos^2r_y\cosec^2d>0.
 \tag{A5}
\end{equation}

\newpage
\section{Non-diagonal responses along the principal directions.}
There are a total of four distinct responses along the principal directions in Eq. 5, where the diagonal responses are accounted for in detail in Sect. III. The remaining two responses are non-diagonal (hall responses) and are distinguished with respect to the direction of the loop separation, i.e. transverse hall response  $J_z(A\hat{x})$ and longitudinal transverse hall response $J_y(A\hat{x})$. As we will show, these responses are specific to the chosen form of the Hamiltonian and are not related to the topological linkage of the nodal loops.

When we set the intermediate parameter $\xi=\frac{\cos r_y-1}{\cos r_x-1}\frac{1}{\sin d}-1$ in the Hamiltonian (Eq. 3) as 1, then the functions $h(\mathbf{k})$ and $g(\mathbf{k})$ (Eqs. 4a,b) are even and odd under the inversion $\mathbf{k}=(k_x,k_y,k_z)\rightarrow(-k_x,k_y,-k_z)$. This symmetry results in a vanishing transverse hall response for all values of impulse $A$. In another words, there exists a critical value of $d$,  $d_{\text{crit}}$, that satisfy $\xi=1$ and is given as
\begin{equation}
d_{\text{crit}}=\sin^{-1}\frac{\cos r_y-1}{2(\cos r_x-1)}\tag{A6}
 \end{equation}
When such solutions for $d_{\text{crit}}$ exist, then the sign of the response will be $\sgn(d-d_{\text{crit}})$.  This remains true even if the nodal loops are no longer linked, i.e. $d_{\text{crit}}>r_y$. In Fig.~\ref{Hall}b, we considered the simple example of circular nodal loops where $r_y=r_x$, then Eq. A6 gives $d_{\text{crit}}=\frac{\pi}{6}$. In the particular example where $\mu=0.1$ and $r_x=r_y=0.6>d_{\text{crit}}=\frac{\pi}{6}$, the nodal tubes are sufficiently thick to have the nodal loops of the Hopf link to touch each other, giving an accidental nodal chain instead. As we can see, the topological linkage of the nodal loops are not important in this response.

The longitudinal transverse Hall response increases in magnitude with $d$ and attains the largest possible magnitude when the nodal loops touch each other. This is illustrated in Fig.~\ref{Hall}b, where we again consider $r_y=r_x=0.6$. We see that when $d=0.2,0.3<r_y$, the response of the Hopf link is significantly weaker, but non-vanishing at small fields. This response grows with $d$ and is the largest when the nodal loops touch each other $d=0.6,0.7\sim r_y$. As for the unlinked case, this response is weaker than that of the nodal chain and stronger than the Hopf link. The response thus has no significant contrast between the topologically linked case and the trivial case, but rather it varies smoothly with $d$.

\section{Nonlinear and anisotropic response of the Hopf link in terms of HHG}
For an oscillatory electric field of frequency $\Omega$, the vector potential amplitude $A=E/\Omega$ is related to the external field amplitude $E$. To quantify the non-linearity of the response expansion, we can do a vectorial Taylor expansion for the response about the point $\mathbf{A_0}$ up to second order:
\begin{align}
&J_i(\mathbf{A})-J_i(\mathbf{A_0})=\sum_{j=x,y,z}\frac{\partial J_i}{\partial A_j}\bigg|_{\mathbf{A_0}}(\mathbf{A}-\mathbf{A_0})\cdot\hat{j}+\nonumber\\&\frac{1}{2}\sum_{j,k=x,y,z}\frac{\partial^2J_i}{\partial A_j\partial A_k}\bigg|_{\mathbf{A_0}}(\mathbf{A}-\mathbf{A_0})\cdot\hat{j}~(\mathbf{A}-\mathbf{A_0})\cdot\hat{k}\nonumber\\&=\sum_{j=x,y,z}a_{ij}\Delta\mathbf{A}\cdot\hat{j}+\frac{1}{2}\sum_{j,k=x,y,z}a_{ijk}\Delta\mathbf{A}\cdot\hat{j}~\Delta\mathbf{A}\cdot\hat{k}\tag{A7}
\end{align}
where we defined the expansion coefficients as the partial derivatives in the expansion, i.e. 
\begin{equation}
a_{ij}=\frac{\partial J_i}{\partial A_j}\bigg|_{\mathbf{A_0}},\quad a_{ijk}=\frac{\partial^2J_i}{\partial A_j\partial A_k}\bigg|_{\mathbf{A_0}}\tag{A8}
\end{equation}
which are approximated using the finite difference method for a small perturbation $|\Delta\mathbf{A}|=|\mathbf{A}-\mathbf{A_0}|=0.005$. These coefficients are related to the conductivities/susceptibilities familiar in optical materials. To understand the evolution of the response surfaces with the field strength $A_0$, we are interested in the expansion coefficients. In particular, for a given field strength $A_0$, the coefficients $a_{ijk}$ quantify the nonlinearity of the responses. Given a direction $\mathbf{A_0}$, there are 18 unique $a_{ijk}$ coefficients (the mixed partial derivatives are symmetrical). 

As an example, we compute the coefficients of the response surfaces (Fig. \ref{A1}(b-d)) for the directions $\mathbf{A_0}=(1,0,1)$, $(-1,0,1)$, $(0,1,0)$, which are displayed in Table I.  Unfortunately, these coefficients alone display limited information since they are local properties of the response manifold and only contain geometric information around their chosen field directions and magnitudes. Combined with our response surface picture, the topology and geometry of the overall response behavior is given a clearer global picture. 

\begin{figure*}
\includegraphics{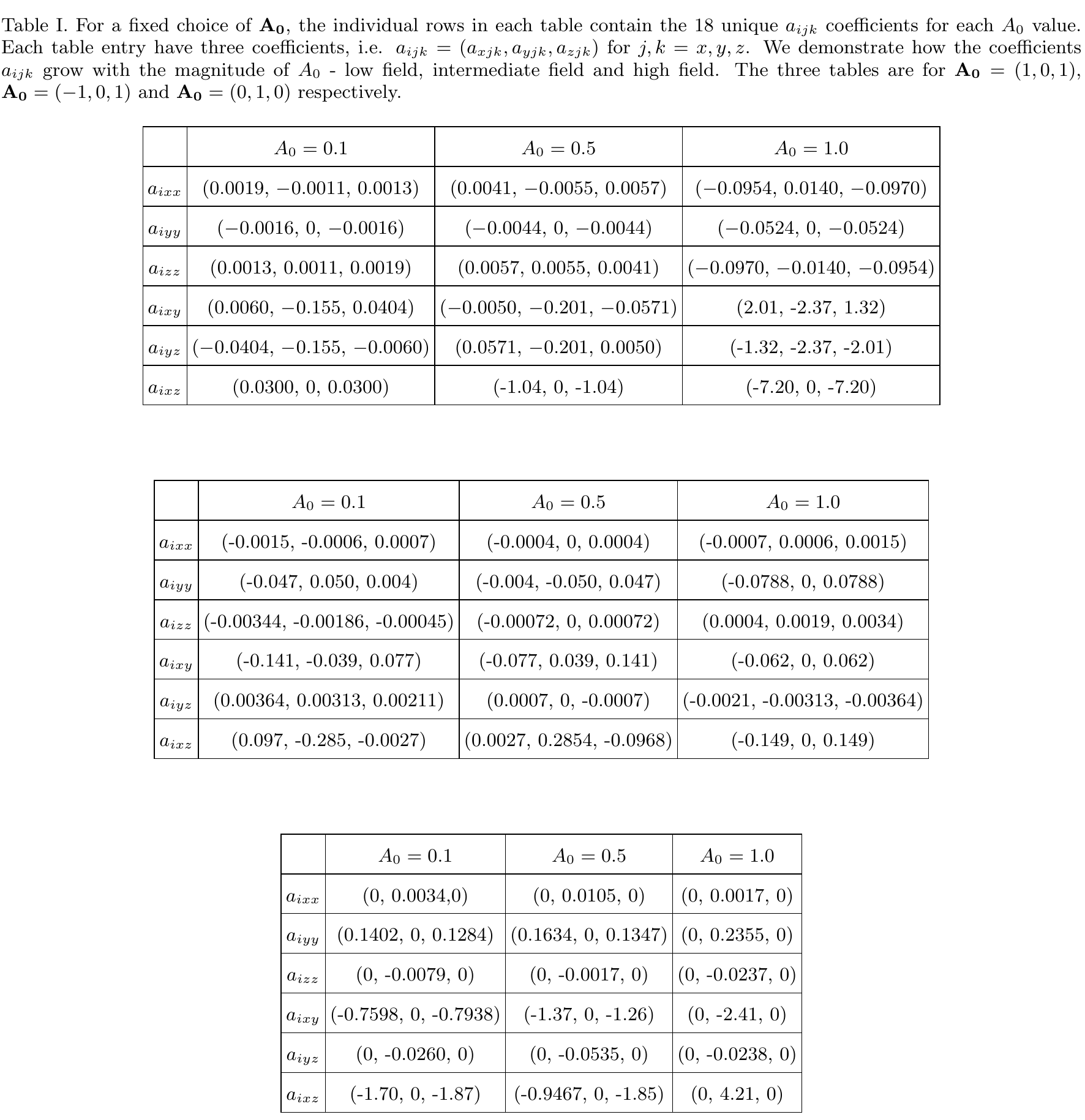}
\end{figure*}

\begin{figure*}
\includegraphics{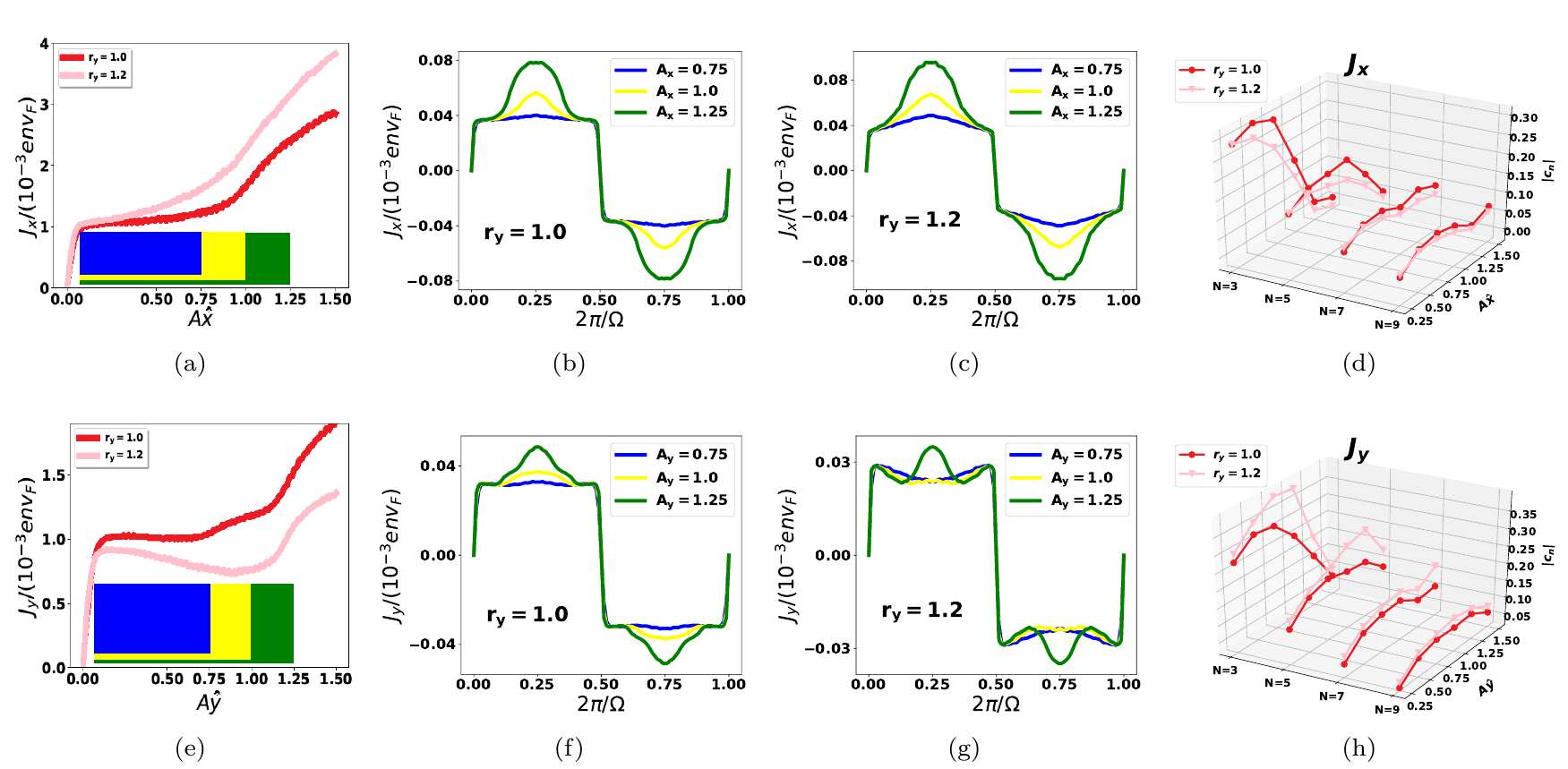}
\caption{To demonstrate the anisotropic response of the Hopf link, we compare its transverse and longitudinal diagonal responses for the parameters $r_x=1$, $d=0.6<r_y=1.0,1.2$. (b,c,f,g)  demonstrates how the response curves distort sinusoidal signals of different amplitudes $A=0.75,1.0,1.25$ near the turning point of the responses (as indicated by the corresponding colored regions in the response curves) for $r_y=1.0$ and $r_y=1.2$ respectively. (a-d) We compare the extent of nonlinearity of the transverse diagonal response $J_x~(A\hat{x})$ for $r_y=1.0,1.2$. In (a), the response for $r_y=1.0$ is clearly more nonlinear. (b,c) The sinusoidal signal for $r_y=1.0$ experiences slightly more distortion. The greater nonlinearity is confirmed in (d) where the HHG coefficients $|c_n|$ are consistently larger for $r_y=1.0$. (e-h) Similarly, we compare the extent of nonlinearity of the longitudinal diagonal response $J_y~(A\hat{y})$. In (e), the response for $r_y=1.2$  is instead more nonlinear. (f,g) The sinusoidal signal for $r_y=1.2$ acquires additional fluctuations of higher frequency. This is confirmed in (h) where the HHG coefficients $|c_n|$ are drastically larger for $r_y=1.2$, especially at the turning point $A_y~1.0$. }
   \label{Appendix: HHG1}
\end{figure*}

To understand the coefficients $a_{ijk}$, we consider the simple case of $j=k$. Consider $\mathbf{A}$ along a particular direction (in general, higher order response coefficients involve more than one component of $\mathbf{A}$), e.g. the principal directions, so we may simply fit the optical response $J$ with a polynomial in $A$:
\begin{equation}
J_i(\mathbf{A}\cdot\hat{j})=a_{i,j}^{(0)}+a_{i,j}^{(1)}A+a_{i,j}^{(2)}A^2+a_{i,j}^{(3)}A^3+\dots\tag{A9}
\end{equation}
The coefficients $a^{(k>1)}_{i,j}$ quantify the  response's nonlinearity. Take the example $r_x=1$, $d=0.6<r_y=1.0,1.2$  (diagonal response curves given in Fig.~\ref{Appendix: HHG1}(a,e)), the response curves are linear in the small field regime $0<A<0.1$ and cubic in the intermediate field regime $0.1<A<1.2$. For the cubic regime, the coefficients are
\begin{equation}
a_{x,x}^{(1)}=0.6935,\quad a_{x,x}^{(2)}=-0.90531,\quad a_{x,x}^{(3)}=1.47878\tag{A10a}
\end{equation}
\begin{equation}
a_{y,y}^{(1)}=0.75134,\quad a_{y,y}^{(2)}=-2.13035,\quad a_{y,y}^{(3)}=1.31172\tag{A10b}
\end{equation}
The drastic difference in $a^{(2)}$'s shows that the longitudinal diagonal response $J_y(A\hat{y})$ is significantly more nonlinear than the transverse diagonal response $J_x(A\hat{x})$, hence the optical response is anisotropic. Another way to quantify the non-linearity of the response is the higher harmonic generation (HHG) coefficients $c_n\sim|J(n\Omega)|/|J(\Omega)|$, i.e. the greater the nonlinearity, the larger the HHG coefficients, which is experimentally measured by the extent of distortion for a sinusoidal signal.
\begin{equation}
J_i(A_j(t))=J_i(A_{j,0}\sin\Omega t)\propto\sin\Omega t+\sum_{n>1}c_n\sin\Omega t\tag{A11}
\end{equation}

\begin{figure*}
\includegraphics{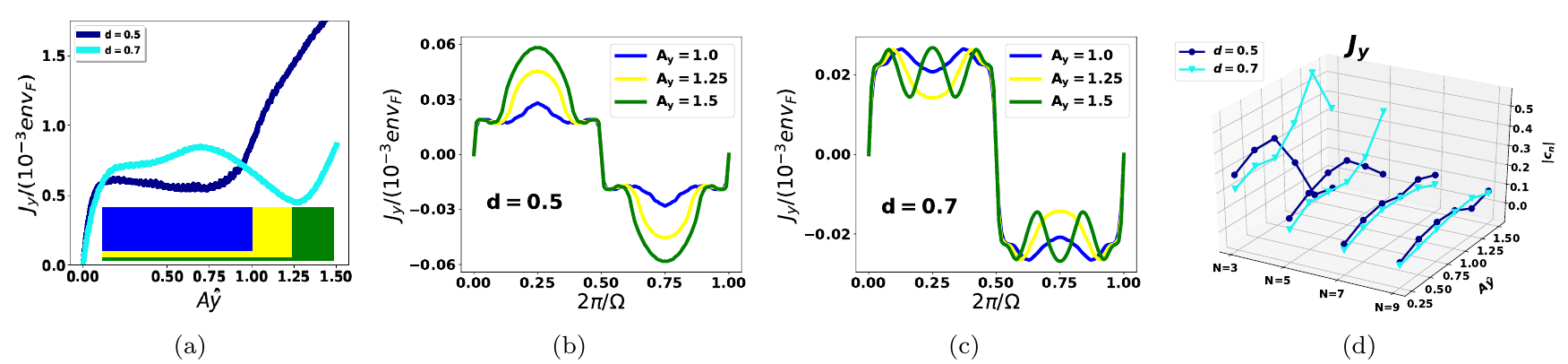}
\caption{To demonstrate the enhanced nonlinearity of the longitudinal diagonal response of the Hopf link, we compare $d=0.5,0.7$ for the parameters $r_x=0.6$, $r_y=1.0>d$. In (a), the response for $d=0.7$ is clearly more nonlinear. (b,c) demonstrates how the response curves distort sinusoidal signals of different amplitudes $A=1.0,1.25,1.5$ near the turning point of the responses (as indicated by the corresponding colored regions in the response curves) for $d=0.5$ and $d=0.7$ respectively.  The sinusoidal signal for the latter acquires additional fluctuations of higher frequency and thus appears significantly distorted. The greater nonlinearity for $d=0.7$ is confirmed in (d) where the HHG coefficients $|c_n|$ are drastically higher for $d=0.7$ at the turning points of the response curves.}
   \label{Appendix: HHG2}
\end{figure*}

For the same example, we compare the HHG coefficients of the transverse and longitudinal diagonal responses. $r_y=1.0$ indeed shows smaller and larger HHG coefficients respectively Fig.~\ref{Appendix: HHG1}(d,h)) in the longitudinal and the transverse directions. The resulting distortion to an arbitrary sinusoidal signal is more apparent in the former. Consider another example, increasing the loop separation $d$ greatly increase the nonlinearity of the longitudinal diagonal response curve and thus lead to enhanced HHG (Fig. \ref{Appendix: HHG2}). This is manifested as a larger kink in the response curves, and thus significant more distortion to the sinusoidal signal.

\begin{figure}
\includegraphics{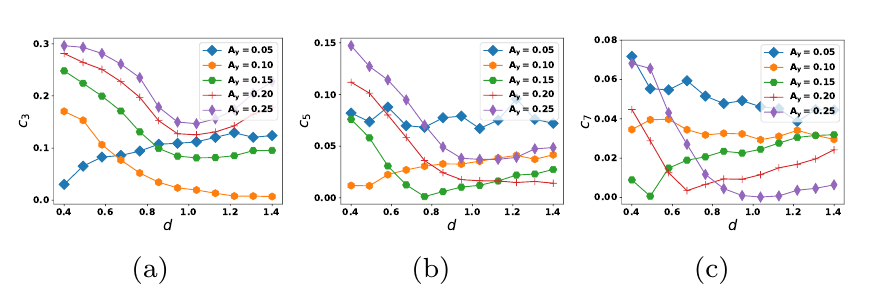}
\caption{For the parameters $r_x=0.9$, $r_y=1.1$ and $\mu=0.1$, we plot the HHG coefficients $|c_n|$ against the loop separation $d$ for various small longitudinal fields $A_y$, and $n=3,5,7$. The HHG coefficients are consistently higher in the $d<r_y$ regime, i.e. topologically linked.}
   \label{Appendix: HHG3}
\end{figure}

Finally, since the longitudinal diagonal response is distinctly more nonlinear when the loops are topologically linked, we thus see enhanced HHG in the topologically linked regime (Fig.~\ref{Appendix: HHG3}). This is consistent with Ref. \cite{PhysRevB.102.035138} which demonstrates that real materials with topologically linked nodal loops exhibit higher HHG compared to materials with only a single nodal loop.

\begin{figure*}
\includegraphics{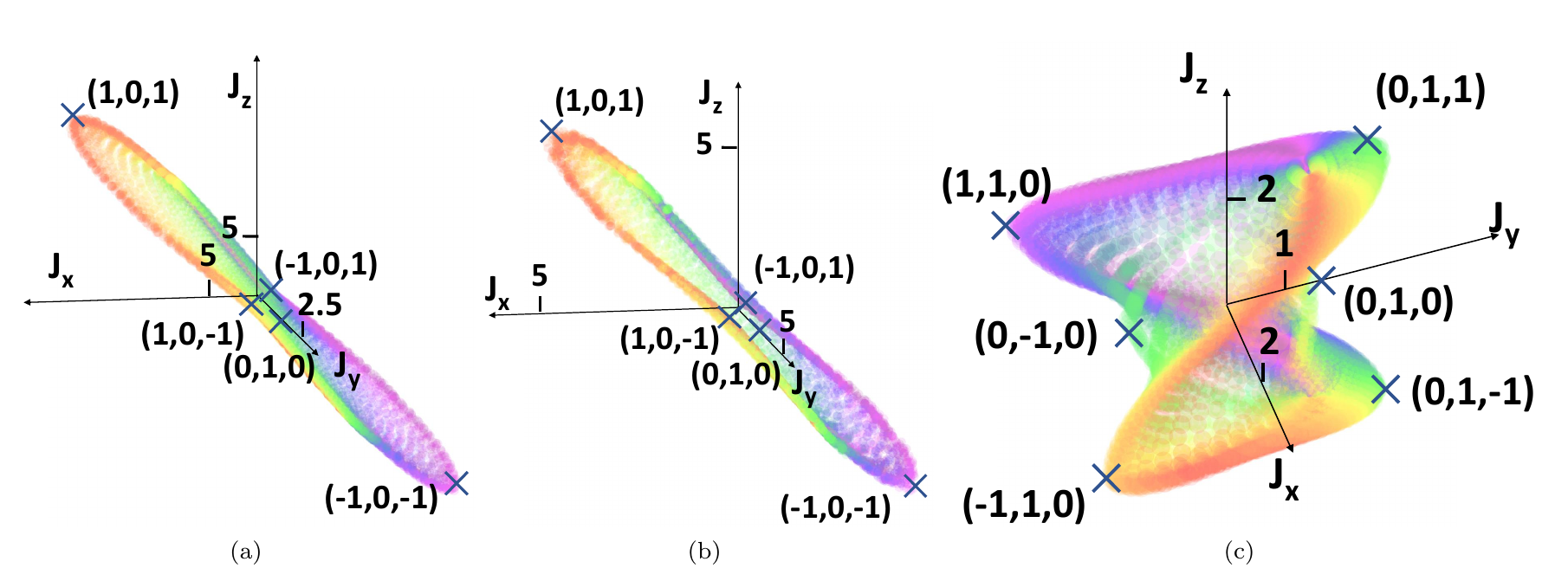}
\caption{For three particular surfaces Figs.~\ref{A2}g,k and Fig.~\ref{A1}g, we present a viewpoint that is different from the previous standardized orientation. This new viewpoint was chosen to highlight the key features of these response surfaces, namely (a-b) the concavity along the directions $\hat{A}=\pm(1,0,-1)$ which characterize its red blood cell-like appearance, as well as, (c) the overall saddle shape appearance.}
   \label{Appendix: diff}
\end{figure*}
\begin{figure*}
\includegraphics{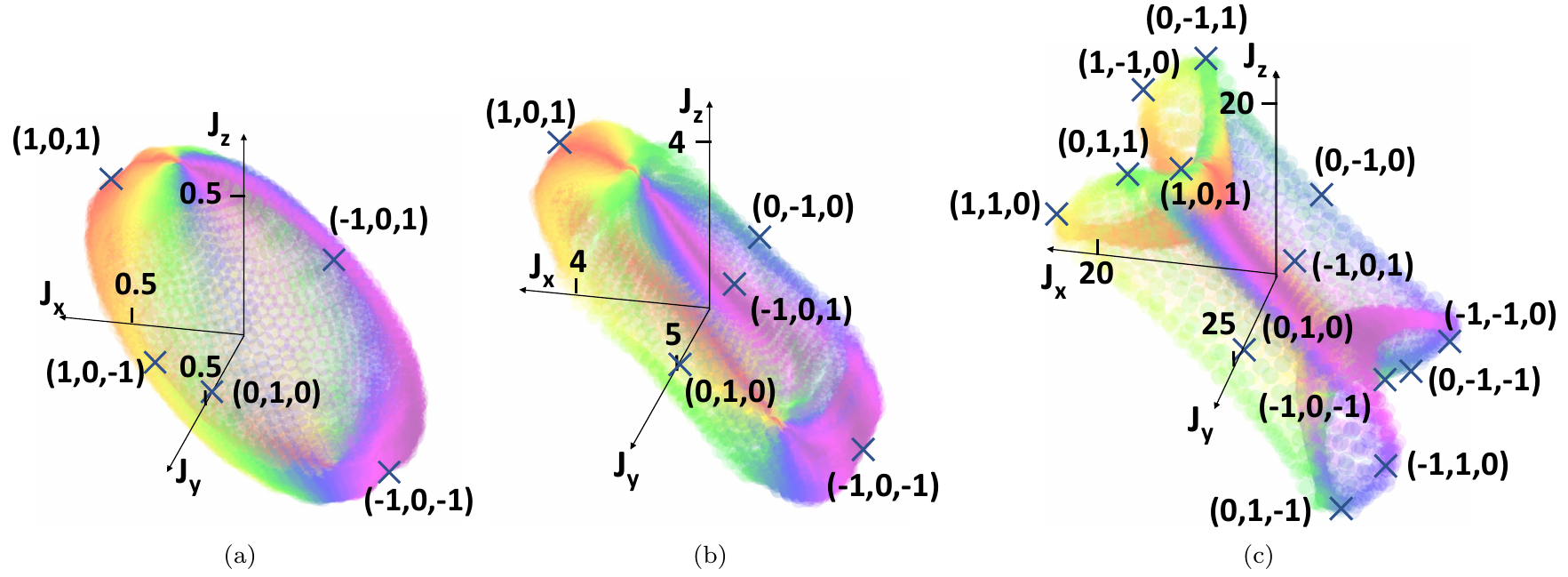}
\caption{For three particular surfaces Figs.~\ref{A2}g,k and Fig.~\ref{A1}g, we present a viewpoint that is different from the previous standardized orientation. This new viewpoint was chosen to highlight the key features of these response surfaces, namely (a-b) the concavity along the directions $\hat{A}=\pm(1,0,-1)$ which characterize its red blood cell-like appearance, as well as, (c) the overall saddle shape appearance.}
   \label{Appendix: similar}
\end{figure*}

 \section{Detailed study of the anisotropy}
In Sect. IV, we demonstrated the evolution of the response surface with field strength. This encapsulates the full information about the response anisotropy and non-linearity, and to some extent the nodal structure and its dispersion. In Figs.~\ref{A1},\ref{A2}, we have illustrated the surface using a common viewpoint to demonstrate this evolution with field strength. But yet, this is not always the best orientation to understand the important characteristics of these surfaces. Locally suppressed response when viewed away from the observer will not be seen in this standard orientation.

For instance in Figs.~\ref{A2}g,k, it is not immediately obvious that these surfaces are indeed reminiscent of the shape of a red blood cell. Here, we better illustrate the distinctive concave shapes that characterize a red blood cell appearance by explicitly demonstrating the concavity along the directions $\hat{A}=\pm(1,0,-1)$. In addition, we clearly highlight the saddle-shaped appearance (which is not made obvious in Fig.~\ref{A1}g) by choosing a better orientation. Here, we can more clearly see that the responses along the directions $\hat{A}=(0,1,1),(1,1,0),(0,1,-1),(-1,1,0)$ are indeed asymmetric as they demonstrate a stronger growth along a preferred sense.

\begin{figure*}
\includegraphics{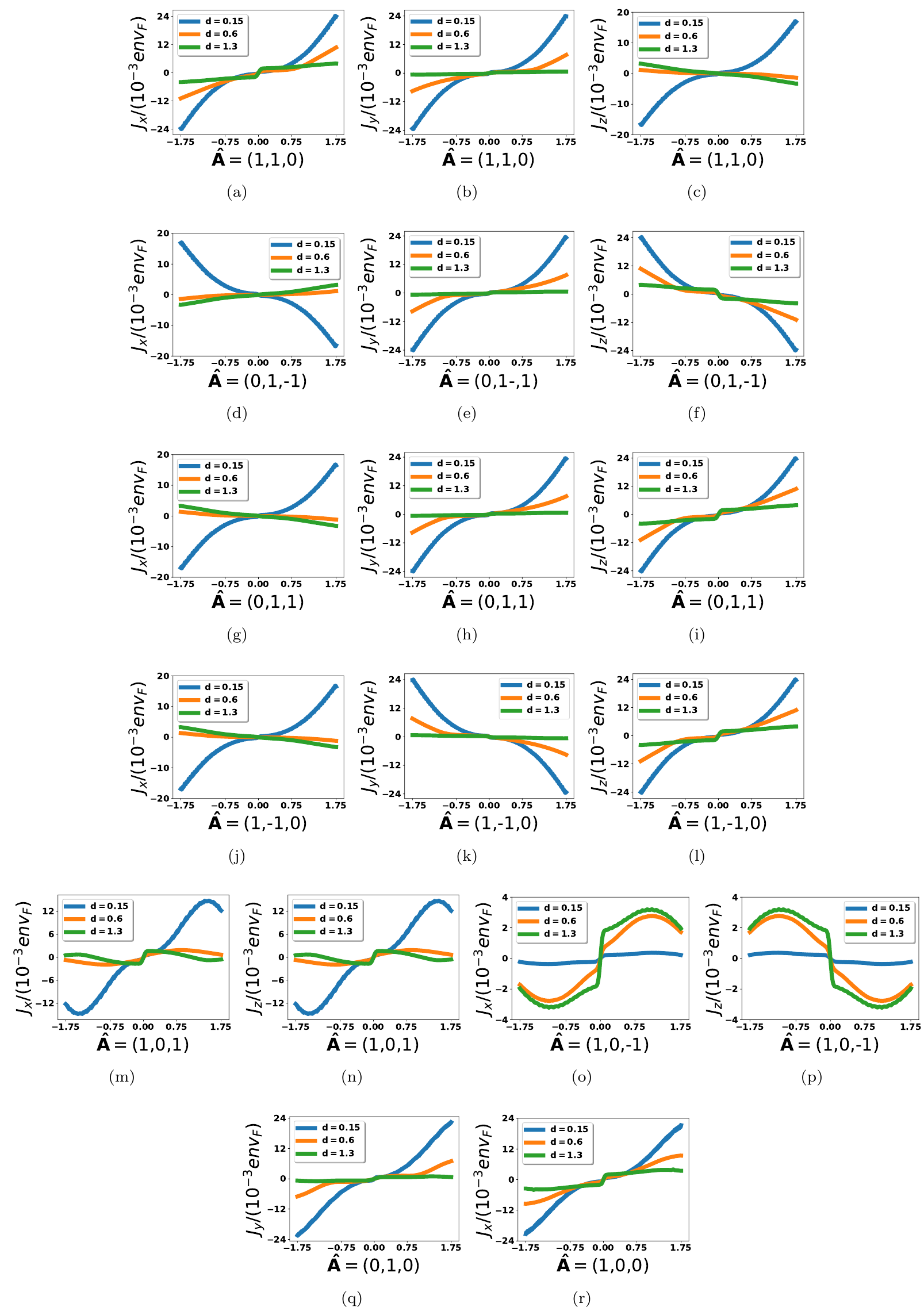}
\caption{The evolution with field strength of the diagonal responses, as well as, the responses along the high symmetry directions that fully characterizes the response surfaces in Fig.~\ref{A1} for the Hopf link $d=0.15<r_y$, the nodal chain $d=0.6=r_y$ and the unlinked case $d=1.3>r_y$. }
    \label{Appendix: ry06rx06response}
\end{figure*}

\begin{figure*}
\includegraphics{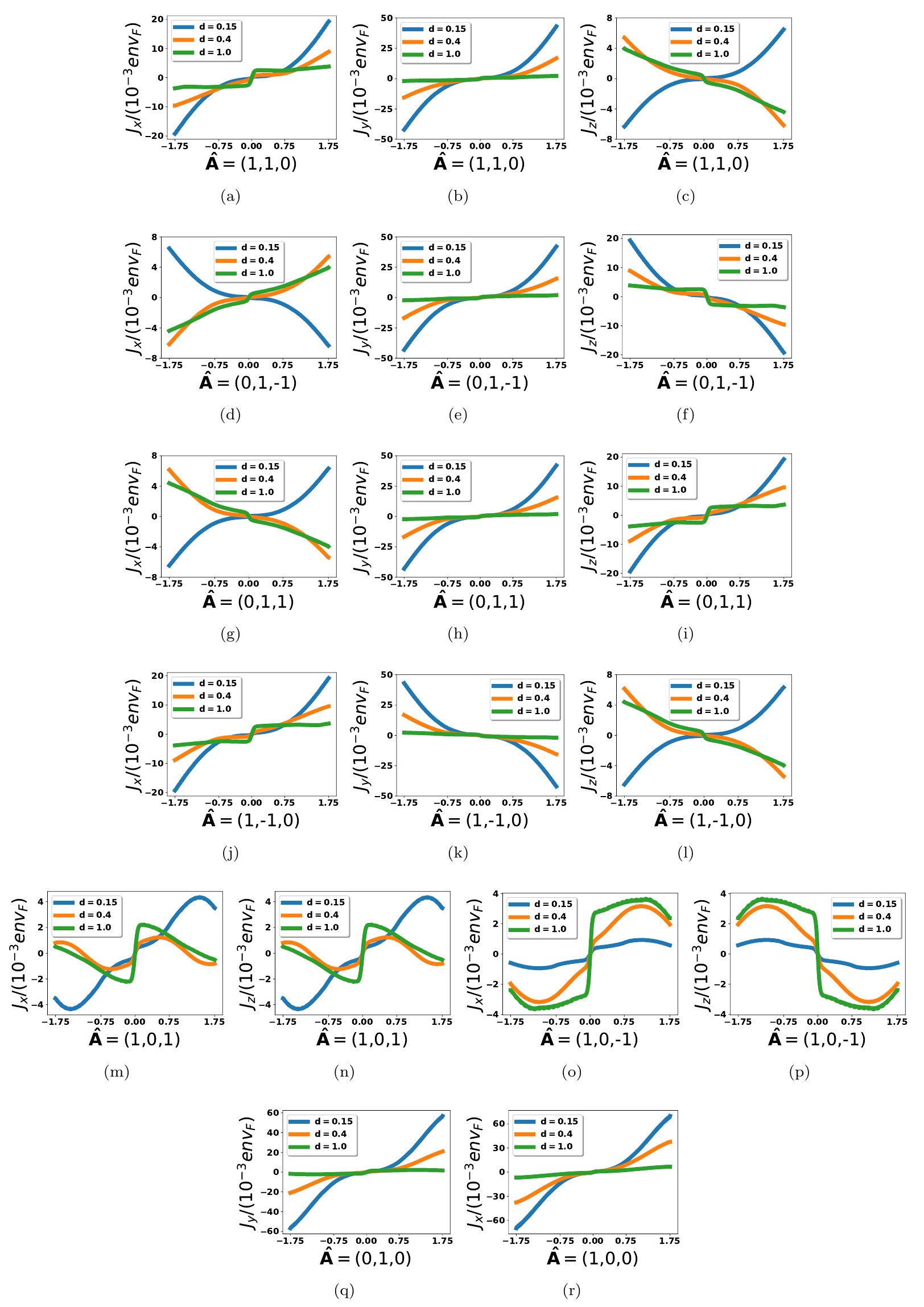}
\caption{The evolution with field strength of the diagonal responses, as well as, the responses along the high symmetry directions that fully characterizes the response surfaces in Figs.~\ref{A2}b-d. Although we have only shown the response surfaces for the Hopf link in Fig.~\ref{A2}, we have also included the responses for the nodal chain case $d=0.6=r_y$ and the unlinked case $d=1.2>r_y$.}
    \label{Appendix: ry06rx04response}
\end{figure*}

\begin{figure*}
\includegraphics{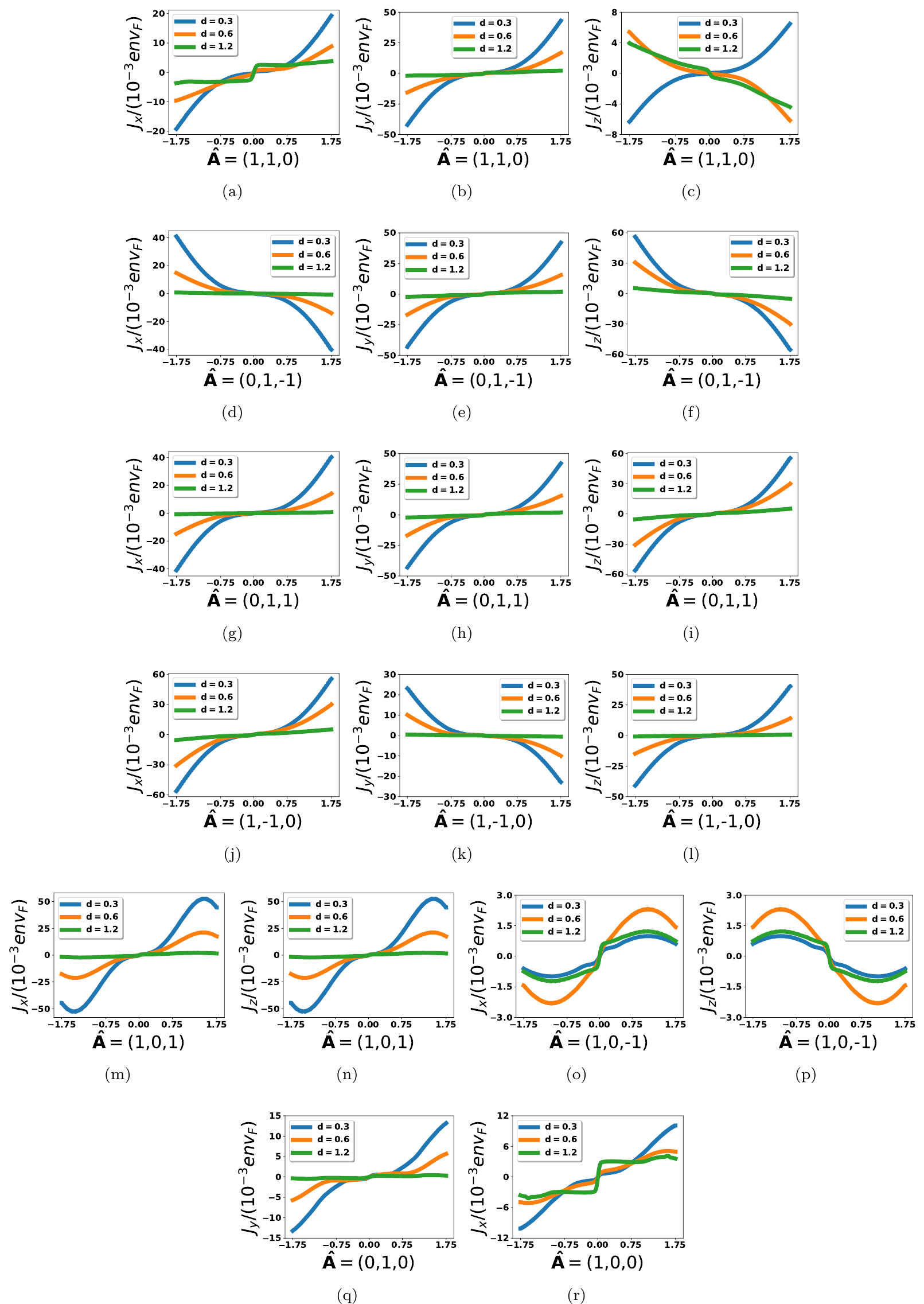}
\caption{The evolution with field strength of the diagonal responses, as well as, the responses along the high symmetry directions that fully characterizes the response surfaces in Figs.~\ref{A2}j-l. Although we have only shown the response surfaces for the Hopf link in Fig.~\ref{A2}, we have also included the responses for the nodal chain case $d=0.4=r_y$ and the unlinked case $d=1.0>r_y$.}
    \label{Appendix: ry04rx06response}
\end{figure*}

Finally, only snapshots of the response surfaces at particular $A_0$ values were chosen in Figs.~\ref{A1},\ref{A2}. This, however, does not capture the full evolution of these surfaces with field strength. Since the diagonal responses and the responses along the high symmetry directions (namely $\hat{A}=(1,\pm1,0),(0,1,\pm1),(1,0,\pm1)$) were shown to be useful in identifying key features of the response surface, we thus show these individual response curves (Fig.~\ref{Appendix: ry06rx06response},\ref{Appendix: ry06rx04response},\ref{Appendix: ry04rx06response}) for the corresponding surfaces. The relative rates of growths of these individual responses can account for the evolution of these surfaces.
\newpage

\newpage
\newpage

\end{document}